\documentclass[prb,10pt,twocolumn,superscriptaddress]{revtex4-1}
\usepackage{amsmath}
\usepackage{latexsym}
\usepackage{amssymb}
\usepackage{graphics,epstopdf}
\usepackage{gensymb}
\usepackage{graphicx}
\usepackage{comment}
\usepackage[colorlinks, linkcolor=red,citecolor=blue,urlcolor=blue]{hyperref}
\newcommand{\be}{\begin{equation}}
\newcommand{\ee}{\end{equation}}
\newcommand{\bea}{\begin{eqnarray}}
\newcommand{\eea}{\end{eqnarray}}
\newcommand{\ket}[1]{\vert #1 \rangle}

\begin{document}
\title{Thermodynamic uncertainty relation for energy transport in  transient regime --- Model study}  % Observation/measurement 
\author{Sushant Saryal}
\affiliation{Department of Physics,
	Indian Institute of Science Education and Research, Pune 411008, India}
\author{Onkar Sadekar}
\affiliation{Department of Physics,
	Indian Institute of Science Education and Research, Pune 411008, India}
	
\author{Bijay Kumar Agarwalla}%	
	\email{bijay@iiserpune.ac.in }

\affiliation{Department of Physics,
		Indian Institute of Science Education and Research, Pune 411008, India}

\date{\today}
		
\begin{abstract}
We investigate transient version of the recently discovered thermodynamic uncertainty relation (TUR) which provides a precision-cost trade-off relation for certain out-of-equilibrium thermodynamic observables in terms of net  entropy production. We explore this relation in the context of energy transport in a bipartite setting for three exactly solvable toy model systems (two coupled harmonic oscillators, two coupled qubits and  a  hybrid coupled oscillator-qubit system) and analyze the role played by the underlying statistics of the transport carriers in TUR. Interestingly, for all these models, depending on the statistics, the TUR ratio can be expressed as a sum or a difference of an universal term which is always greater or equal to 2 and a corresponding entropy production term. We find that the generalized version of the TUR, originating from the universal fluctuation symmetry is always satisfied. However, interestingly, the specialized TUR, a tighter bound, is always satisfied for the coupled harmonic oscillator system obeying  Bose-Einstein statistics. Whereas, for both the coupled qubit, obeying Fermi-like statistics and the hybrid qubit-oscillator system with mixed Fermi-Bose statistics, violation of tighter bound is observed in certain parameter regimes. We have provided conditions for such violations. We also provide a rigorous proof following the non-equilibrium Green's function approach that the tighter bound is always satisfied in the weak-coupling regime for generic bipartite systems.
\end{abstract}

\maketitle 

Small scale systems are prone to fluctuations \cite{Ritort}. Characterizing and quantifying thermal and quantum fluctuations for small scale systems are therefore important both from the fundamental and as well as practical point-of-view \cite{spin-heat, heat-fluc}. Last two decades have seen a plethora of interesting works in this direction. In particular, the discovery of non-equilibrium universal fluctuation relations \cite{fluc-old1, fluc-old2, fluc-old2a, fluc-old3, fluc-old4, JarzW, fluct1, fluct2, fluct3, SaitoUts,campisi-measurement}, concerning with microscopic description of systems, have provided a deeper understanding about nonequilibrium thermodynamics and have greatly contributed in establishing the rapidly growing field of stochastic and quantum thermodynamics \cite{st-thermo1,st-thermo2, QT1, QT2}. 

Along this direction, very recently, an interesting trade-off relation involving relative fluctuations of certain non-equilibrium observables has been put forward, providing a lower bound on these fluctuations in terms of the associated entropy production. Various versions of this relation, now collective referred to as the '{\it Thermodynamic Uncertainty relations}' (TUR's) have been proposed and furthermore its generality has been examined in many different contexts: such as for, steady state systems following classical Markovian dynamics, periodically driven systems, quantum transport problems, molecular motors, finite-time statistics, first-passage times,  etc
\cite{Barato:2015:UncRel,Gingrich:2016:TUP,Horowitz:2017:TUR,Pietzonka:2016:Bound, Proesmans:2017:TUR,Dechant:2018:TUR,Pietzonka:2017:FiniteTUR,Horowitz:2017:TUR,Pigolotti:TURF, Dechant:2018:TUR,Dechant:2018:TUR,Gingrich:2017,Hasegawa1,TUR-gupta,Hyeon:2017:TUR,Koyuk:2018:PeriodicTUR,Gabri,Dechant:2018:TUR,Hwang, Oren, Falasco,Vu, Hasegawa2, Sasa:TUR, Baiesi, Bio,Interacting,Mayank,Van,SamuelssonM,Udo:TURB,Saito,Garrahan18,SamuelssonM,Hyst,Garrahan:2017:TUR,Passage,BijayTUR,TUR-bijay1,TURQ,SamuelssonQP,JunjieTUR, Goold,Polettini:2016:TUP,Landi-PRL,Isometric, Horowitz:2019:TUR,TUR-driving-udo, TUR-opensys,TUR-Otto,tur_energy,tur_superconductor,tur_turing,tur_hyper,tur_continuous}. Parallel to these theoretical developments, experimental studies of TUR relations for classical and quantum systems have also been reported very recently \cite{Oren, Soham_TUR,tur_kproof,tur_infoengine}.

Here we are  interested in understanding the transient version of the TUR relation in the context of energy exchange that takes place between two quantum systems which are initially equilibrated at different temperatures.  For such transport, a non-universal tighter version of the TUR bound (T-TUR) is given as
\be
\frac{\langle Q^2 \rangle_c}{\langle Q \rangle^2} \geq \frac{2}{\langle \Sigma \rangle},
\label{eq:fund-TUR}
\ee
where $Q$, a stochastic variable, is the integrated energy current over a certain time duration. $\langle Q \rangle, \langle Q^2 \rangle_c$ represents the average  energy exchange and the corresponding noise, respectively. $\langle \Sigma \rangle \geq 0 $ represents the average entropy production in the energy exchange process and further characterizes how far the composite system is driven away from the initial condition.

A loose but a generalized version of the bound (G-TUR1) compared to Eq.~(\ref{eq:fund-TUR}) \cite{Vu} was recently derived following the fundamental energy exchange fluctuation relation (XFT) \cite{JarzW} where the RHS of Eq.~(\ref{eq:fund-TUR}) was modified to 
\be
\frac{\langle Q^2 \rangle_c}{\langle Q \rangle^2} \geq \frac{2}{\exp{\langle \Sigma \rangle} -1}.
\label{eq:fund-TUR-1}
\ee
In fact, a more tighter version (still loose compared to Eq.~(\ref{eq:fund-TUR})) of the generalized bound in Eq.~(\ref{eq:fund-TUR-1}) was obtained by Timpanaro {\it et al.} \cite{Landi-PRL}(G-TUR2), given as 
\be
 \frac{\langle Q^2 \rangle_c}{\langle Q \rangle^2} \geq f(\langle \Sigma \rangle),
\label{eq:fund-TUR-2}
\ee
where $f(x)={\rm csch}^2(g(x/2))$ and $g(x)$ is the inverse function of $x \tanh(x)$. 

Of-course, it is clear that, systems satisfying XFT will follow the G-TUR1 and G-TUR2. However, it is still an interesting question to ask under what conditions the tighter version i.e., the T-TUR bound in Eq.~(\ref{eq:fund-TUR}) will be preserved. Very recently, the usefulness of the T-TUR bound was proposed to infer the net entropy production for complex non-equilibrium systems \cite{Supriya}. 

In this work, we  analyze TUR bounds for quantum energy transport by focusing on three different model systems characterized by different quantum statistics: bosonic, fermionic and hybrid Fermi-Bose statistics for the transport carriers. Since it is well known that quantum statistics plays a key role in the transport properties, we ask how does it effect the transient TUR bounds? Interestingly, we find that when energy exchange takes place between two simple quantum harmonic oscillators, obeying Bose-Einstein statistics, the T-TUR in Eq.~(\ref{eq:fund-TUR}) is always satisfied. Whereas, in the other extreme scenario, i.e., when each system consists of a single qubit, following Fermi-like statistics, violation for the T-TUR is observed in certain parameter regimes. As a final interesting example, we consider a hybrid setup consisting of a single quantum harmonic oscillator and a qubit and analyze the impact of hybrid-statistics on TUR. We also show that for general bipartite setup, the T-TUR is always satisfied in the weak-coupling regime. 
Expectedly, in all these setups, the generalized version of TUR is  satisfied due to the underlying XFT for energy exchange. 

The paper is organized as follows: We start in section II with a  brief introduction about obtaining the characteristic function (CF) for the energy exchange following the two-time measurement protocol and describe the associated XFT. In Sec.~III we introduce three toy models, provide derivation for the exact CF's and analyze the corresponding TUR.  We summarize our main findings in Sec.~IV.  We provide certain details including a proof for the T-TUR bound in weak-coupling regime in the appendix.

%--------------------------------------------------------------------------------------
\section{ Energy exchange statistics and the characteristic function}
%--------------------------------------------------------------------------------------
In this section we briefly outline the theory behind obtaining the quantum energy exchange statistics for a generic out-of-equilibrium bipartite setup. Under this setting, one considers two systems with Hamiltonians $H_1$ and $H_2$ that are initially  $(t=0^{-})$ decoupled with composite density matrix given by a product state,
${\rho}(0) = {\rho}_1 \otimes {\rho}_2$,  with ${\rho}_{i} = \exp\big[{-\beta_{i} H_{i}}\big]/{\mathcal Z}_{i}, i=1,2$ being the initial Gibbs thermal state with inverse temperature $\beta_{i}=1/ T_{i}$ (we set $k_B=\hbar =1$) and ${\mathcal Z}_{i}={\rm Tr} \big[e^{-\beta_{i} H_{i}}\big]$ is the corresponding canonical partition function.  To allow energy exchange, an interaction term between the two systems, denoted as $V$, is suddenly switched on at $t=0$ and suddenly switched off after a duration of $t=T$. The composite system in this interval evolves unitarily with the total Hamiltonian $H=H_1+H_2+V.$ 

It is now a well known fact that for quantum systems quantities such as energy current, work, or the associated entropy production are not direct observables but rather depends on the measurements of relevant Hamiltonians at the initial and final time of the process.  Therefore, to construct the probability distribution function (PDF)  \cite{fluct1, fluct2, campisi-measurement} for energy exchange, projective measurements of the system Hamiltonians $H_1$ and  $H_2$ should be carried out simultaneously in the beginning and at the end of the process.
Following this, the joint PDF, $p_{T}(\Delta E_1, \Delta E_2)$, corresponding to the energy change ($\Delta E_i, i=1,2$) of both the systems can be constructed as
\begin{equation}
p_{T}(\Delta E_1, \Delta E_2)\!=\! \sum_{m,n} \Big(\prod_{i=1}^2 \,\delta(\Delta E_i - (\epsilon_m^i -\epsilon_n^i)) \Big) p_{m|n}^T p_{n}^{0},
\label{eq:prob}
\end{equation}
where $p_n^0 = \prod_{i=1}^2 e^{-\beta_i \epsilon_n^i}/{\cal Z}_i$ corresponds to the probability to find the system initially in the common energy eigenstate $|n\rangle=|n_1, n_2\rangle$ of the composite system where $|n_i\rangle$ and $\epsilon_n^i$ are energy eigenstate and eigenvalue respectively of system $i$ after the first projective measurement. The second projective measurement at the final time ($t=T$) leads to the collapse of the state of composite system to another common energy eigenstate $|m\rangle=|m_1,m_2\rangle$. The transition probability between these states is given by $p_{m|n}^T = \langle m |\,{\cal U}(T,0)|n\rangle|^2$ with ${\cal U}(t,0)=e^{- i {H} t}$ being the global unitary propagator with the total Hamiltonian ${H}$. Now one can show that for autonomous and time-reversal invariant quantum systems evolving unitarily $p_{m|n}^T =p_{n|m}^T $. This condition is also known as the principle of micro-reversibility in the literature \cite{fluct1,fluct2}. Using this condition in Eq.~(\ref{eq:prob}) one receives the following universal symmetry for this joint PDF, 
\begin{equation}
p_{T}(\Delta E_1,\Delta E_2) = e^{\beta_1 \Delta E_1 +  \beta_2 \Delta E_2} \,p_{T}(-\Delta E_1, -\Delta E_2).
\label{fluc}
\end{equation}
At this junction, it is important to point out that under general coupling scenario the energy change of an individual system can not be interpreted as heat as part of the energy change may be used in turning on and off the interaction ($V$) between the two systems.  However, in the weak-coupling limit ($V \ll H_{1,2}$), it is safe to interpret this energy change as heat. One can then define heat as $Q = -\Delta E_1 \approx \Delta E_2$ which following Eq.~(\ref{fluc}) then leads to a heat exchange fluctuation relation (XFT) 
\be
 p_{T}(Q) = e^{\Delta\beta Q} p_{T}(-Q),
 \label{eq:XFT}
 \ee
 where $\Delta \beta= \beta_2-\beta_1$. As per our convention, heat flowing out from system 1 to system 2 is considered as positive.

The central object of interest in our work is the characteristic function (CF) corresponding to the PDF for energy exchange which is obtained by performing a Fourier transformation (FT) of the probability distribution:
\begin{eqnarray}
\!\!\!\!\!\!\chi_{T}(u)&\!\!=\!\!&\int dQ \, e^{i u Q} \, p_T(Q) \nonumber \\
=&&\!\!{\rm Tr}\Big[{\cal U}^{\dagger}(T,0) (e^{-i u H_1} \otimes {1}_2)\, {\cal U}(T,0)  (e^{i u H_1} \otimes {1}_2) \rho(0)\Big].\, \,\,\,\,\,\,\,
\label{eq:CF-TTM}
\end{eqnarray}
Here $u$ is a variable conjugate to $Q$.  In terms of CF, the XFT for heat in Eq.~(\ref{eq:XFT}) translates to $\chi_{T}(u) = \chi_{T}\big(-u + i \Delta \beta \big)$ \cite{JarzW,Bijay12, Saito07,Lutz_2018, XFT-agarwalla,XFT-theory}.

It is important to note that, for a special choice of the coupling Hamiltonian $V,$ satisfying the commutation relation $[V, H_1+H_2]=0$, the total internal energy $H_1 + H_2 $ is a constant of motion which imply that the change of energy for one system is exactly compensated by the other one. In other words, there is no energy cost  involved in turning on and off the interaction between the systems. Such type of coupling is known as the thermal coupling.  Therefore, under this symmetry condition the definition for heat $Q = - \Delta E_1 = \Delta E_2$ becomes exact for arbitrary coupling strength.  More generally, the XFT for heat is preserved exactly in this limit (see Appendix A for the details of the proof following the CF of heat).

In what follows, we study three different toy models with different underlying quantum statistics in the thermal coupling limit and thus getting rid of any ambiguity with the definition for heat. We derive exact analytical expressions for the CF and then analyze the impact of quantum statistics on the TUR bounds.

%--------------------- Figure---------------------------------
\begin{figure}
\centering
\includegraphics[width=0.7\columnwidth]{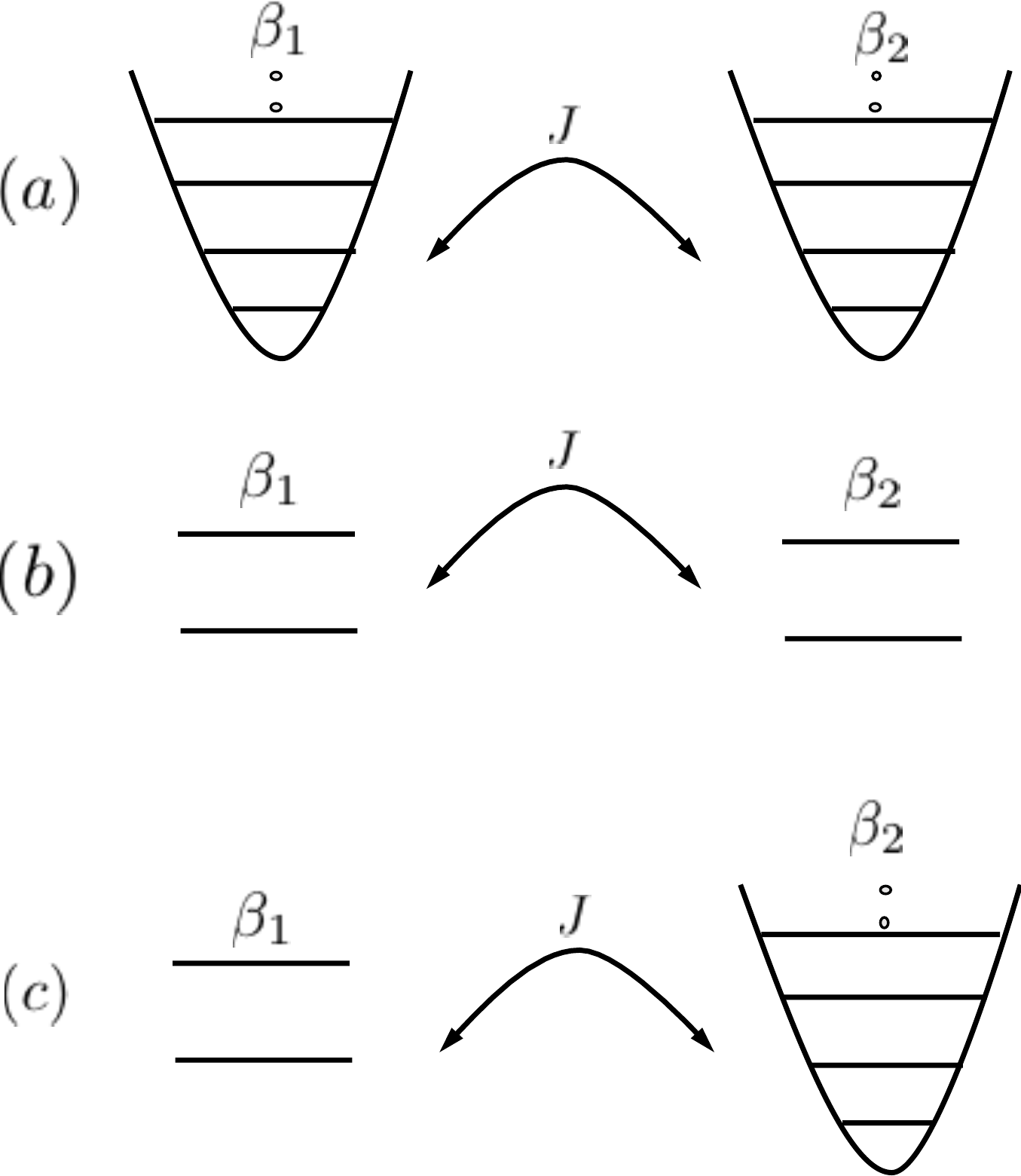}
\caption{Schematic for three different toy models that we investigate in this paper: (a) coupled two-oscillator system, (b) coupled two-qubit system and (c) coupled hybrid qubit-oscillator system. Each system is prepared initially in equilibrium at a particular inverse temperature $\beta_i = 1/k_B T_i$. A finite thermal coupling with coupling strength $J$ allows energy exchange between the systems.}
\label{schematic}
\end{figure}
%--------------------- Figure---------------------------------

%--------------------- Figure---------------------------------
\begin{figure*}
\centering
\includegraphics[scale=1]{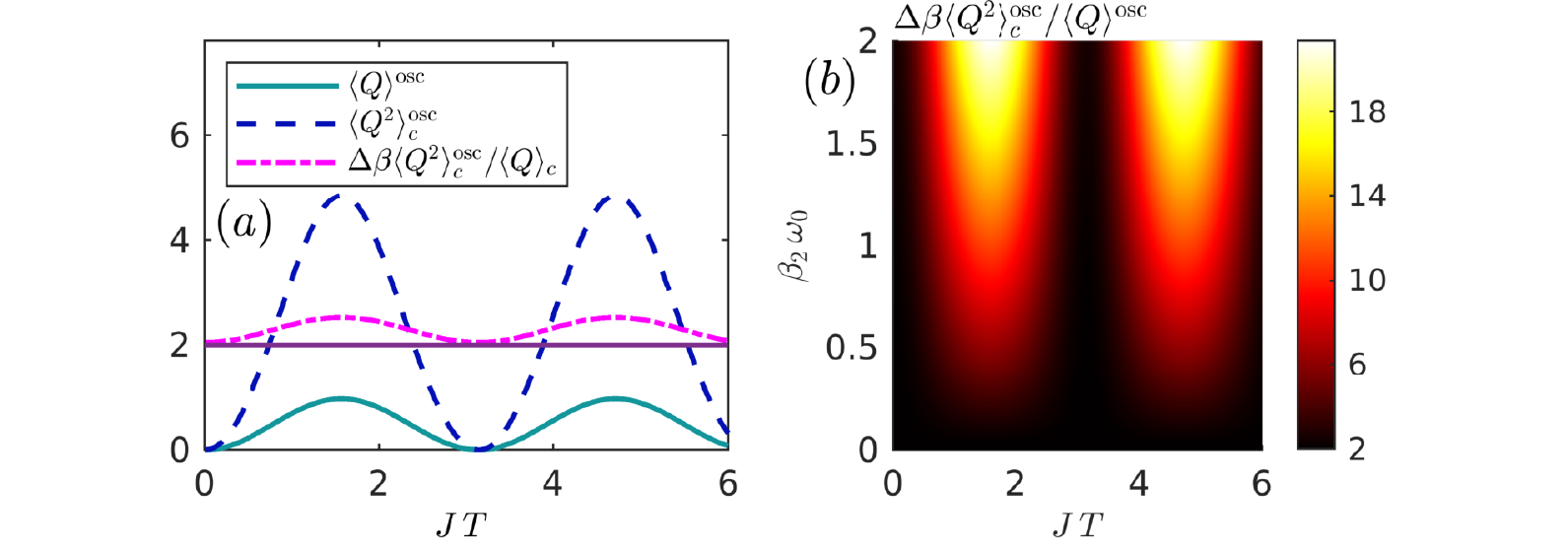} 
\caption{(Color Online)  (a) Plot for average energy change $\langle Q \rangle^{\rm osc}$ (solid), fluctuation $\langle  Q^2  \rangle^{\rm osc}_c$ (dashed) and the corresponding TUR ratio $\Delta \beta  \langle Q^2  \rangle^{\rm osc}_c /\langle Q \rangle^{\rm osc}$ (dashed-dotted) as a function of $J\,T$. For reference a line is drawn at the value 2. The parameters are $\beta_1 \omega_0=0.5,  \beta_2 \omega_0=1$. (b) Two-dimensional plot for TUR ratio ($\Delta \beta \frac{\langle  Q^2  \rangle^{\rm osc}_{c}}{\langle Q \rangle^{\rm osc}}$) as a function of $J\,T$ and $\beta_2 \omega_0$. We set $\beta_1 \omega_0 = 0.1$.}
\label{fig1:HO}
\end{figure*}
%--------------------- Figure---------------------------------

%--------------------- Section III---------------------------------
\section{Models and TUR}
\subsection{Two-oscillator system}
As a first example, we consider a bipartite setup where each system consists of a single quantum harmonic oscillator (see Fig.~(\ref{schematic}a)). The total Hamiltonian is given as
\bea
{H}_{\rm osc} &=&  \omega_0 a^{\dagger}_{1}  a_{1} \otimes 1_2
+ 1_1 \otimes  \omega_0 a^{\dagger}_{2}  a_{2} 
\nonumber\\
&+&  J\,(a_1^{\dagger} \otimes a_2 +  a_2^{\dagger} \otimes  a_1 ).
\label{Ham_osc}
\eea 
where the first two terms ($H_1={\omega_0} \, a^{\dagger}_{1}  a_{1} \otimes 1_2$ and  $H_2= 1_1 \otimes   \omega_0 a^{\dagger}_{2}  a_{2} $ ) correspond to two non-interacting  quantum harmonic oscillators  with  $a_{i} (a_{i}^{\dagger})$ being the bosonic annihilation (creation) operator for the $i$-th oscillator. The last term, we denote here as $V$, represents  a bilinear interaction between the oscillators with coupling strength $J$. Note that the frequency of both the oscillators ($\omega_0$) is chosen to be identical which ensures the thermal coupling condition i.e., $[V, H_1+H_2]=0$.  Recall that, before turning on the interaction $V$, each oscillator is thermalized independently at a particular temperature which can be achieved by placing the system in weak contact with a thermal bath. After that, the oscillators are separated from the bath and the the interaction between them is turned on to allow energy exchange for a certain duration $T$. The corresponding CGF ${\cal G}^{\rm osc}_{T}(u) = \ln \chi^{\rm osc}_{T}(u)$  can be obtained exactly and is given as, 
\begin{eqnarray}
 {\cal G}_{T}^{\rm osc} (u) &&= - \ln \Big[1 \!-\! \sin^2 \Big(J T\Big) \Big\{n_1(\omega_0) \, (1 + n_2(\omega_0)) \big(e^{i u \omega_0} \!-\!1\big) \nonumber \\
&&+ n_2 (\omega_0)(1 \!+\! n_1(\omega_0)) \big(e^{-i u \omega_0} \!-\!1\big)\Big\}\Big],
\label{CGF:osc}
\end{eqnarray}
where $n_{i}(\omega_0)= (e^{\beta_i \omega_0} - 1)^{-1}$, $i=1,2$ is the Bose-Einstein distribution function.  We provide the derivation of Eq.~(\ref{CGF:osc}) in the Appendix C by employing the Keldysh non-equilibrium Green's function (NEGF) approach \cite{NEGF1, NEGF2, NEGF3, NEGF4, Mahan}. Note that, a similar model was previously studied in the context of fluctuation symmetry where the CGF was obtained only in the weak-coupling regime \cite{XFT-Akagawa}. Very recently, this model is studied in the context of quantum heat engines \cite{TUR-Otto}. It is easy to verify that the above CGF expression preserves the XFT for arbitrary $T, J, \beta_1$ and $\beta_2$. 

To analyze the TUR bound, we now get the expressions for the average energy change and the associated noise. These are easily obtained by taking successive derivatives of ${\cal G}_{T}^{\rm osc} (u)$  with respect to $iu$. We receive, (for notational compactness, below we denote $n_{i}(\omega_0)$ as $n_{i}$)
\begin{widetext}
\begin{align}
\langle Q \rangle^{\rm osc} &= \omega_0  {\cal T}_{T}(J) \Big[n_1 \!-\!n_2\Big], 
\label{current}
\\
\langle Q^2 \rangle_{c}^{\rm osc} &= \omega_0^2  \Big[{\cal T}_{T}(J) \Big( n_1 (1\!+\!n_2) \!+\! n_2 (1\!+\!n_1)\Big)\!+\! {\cal T}_{T}^2(J) \,\big(n_1\!-\!n_2\big)^2 \Big],\,\,\,\,\,\,  
\label{fluc}
\end{align}
\end{widetext}
where we define ${\cal T}_{T}(J) = \sin^2 \big(J T\big)$.  Since the second term in Eq.~(\ref{fluc}) is always positive, we receive the following inequality,
\begin{eqnarray}
 \langle Q^2 \rangle ^{\rm osc}_{c} \geq \omega_0^2  {\cal T}_{T}(J) \Big( n_1 (1\!+\!n_2) \!+\! n_2 (1\!+\!n_1)\Big),
\label{ineq: harmonic}
\end{eqnarray}
where the equality sign corresponds to equilibrium situation i.e., $\beta_1=\beta_2$. We now make use of the following important relation involving the Bose-Einstein distribution function,
\begin{align}
n_1(1\!+\! n_2) + n_2 (1\!+\!n_1) &= \coth \frac{\Delta \beta \omega_0}{2} \,  \big(n_1 \!-\! n_2 \big) 
\label{equality-Bose}
\\
& \geq \frac{2}{\Delta \beta \omega_0} \big(n_1 \!-\! n_2 \big),
\label{bose-relation}
\end{align}
where in the second line we have used the inequality $x \coth(x) \geq 1$. Substituting this in Eq.~(\ref{ineq: harmonic}) and using  Eq.~(\ref{current}), it is easy to see that $ \Delta \beta \frac{ \langle Q^2 \rangle^{\rm osc}_{c}}{\langle Q \rangle^{\rm osc}} \geq 2$ which imply that for the coupled quantum harmonic oscillator setup displaying bosonic statistics the T-TUR is always satisfied. 

In fact, an interesting observation can be made by arranging the TUR ratio ($\Delta \beta \frac{ \langle Q^2 \rangle^{\rm osc}_{c}}{\langle Q \rangle^{\rm osc}}$) using the expressions for the cumulants (Eq.~(\ref{current}) and Eq.~(\ref{fluc})) and Eq.~(\ref{equality-Bose}). One receives, 
\be
\Delta \beta \frac{ \langle Q^2 \rangle^{\rm osc}_{c}}{\langle Q \rangle^{\rm osc}} = \Delta \beta \omega_0 \coth \frac{\Delta \beta \omega_0}{2} + \langle \Sigma \rangle^{\rm osc} \geq 2.
\label{TUR-ratio-HO}
\ee
Interestingly, the first term here is independent of the coupling information between the systems and is always greater or equal to 2 (equality holds in equilibrium). In contrast, the second term, is exactly the average entropy production for the oscillator system which along with the temperature difference also importantly depends on the dimensionless coupling $J\,T$.  As the average entropy production remains always positive, $\langle \Sigma \rangle^{\rm osc} \geq 0$, once again we arrive at the same conclusion that the T-TUR for this setup is always satisfied. Also note that the validity T-TUR immediately implies that the GTUR-1 (Eq.~(\ref{eq:fund-TUR-1})) and G-TUR2 (\ref{eq:fund-TUR-2}) are also trivially obeyed.

In Fig.~(\ref{fig1:HO}(a)) we plot the first two cumulants and the corresponding TUR ratio as a function of $J\,T$. Fig.~(\ref{fig1:HO}(b)) corresponds to a two-dimensional plot for the TUR ratio as a function of $J\,T $ and $\beta_2\omega_0$. We set $\beta_1 \omega_0=0.1$ in the simulation. The cumulants as well as the TUR ratio oscillates with  $J\,T$ with periodicity $\pi$. The value for TUR ratio is always larger than 2 and matches with the theoretical prediction. For a fixed value of $J \,T$, the TUR ratio increases monotonically with increasing $\Delta \beta$.

%--------------------- Subsection-Two qubit---------------------------------
\subsection{Two-qubit system}
We next consider another toy model, which we refer here as the XY-model, consisting of two qubits (see Fig.~(\ref{schematic}b)). We write the total Hamiltonian  as
\bea
{H}_{XY} &=& \frac{\omega_0}{2} \sigma_1^z \otimes 1_2 
+ 1_1 \otimes \frac{\omega_0}{2} \sigma_2^z  
\nonumber\\
&+& \frac{J}{2}\,( \sigma_1^x \otimes \sigma_2^y - \sigma_1^y \otimes \sigma_2^x).
\label{eq:htotal1}
\eea
$\sigma_i, i=x,y,z$ are the standard Pauli matrices.  Once again, this model satisfy the thermal coupling condition. Very recently, this model was experimentally realized by some of us in  the nuclear magnetic resonance (NMR) setup  to asses the validity of the transient TUR by obtaining the cumulants of energy exchange following quantum state tomography technique \cite{Soham_TUR}. The same model was also used earlier to examine the XFT by measuring the CF for heat exchange employing the ancilla-based interferometric technique \cite{ancilla-1,ancilla-2,ancilla-3,Bijay-expt}. 
We therefore keep some of our discussion here brief and request the readers to see Ref.~(\onlinecite{Soham_TUR}) for the details about the model. 

One can  analytically compute the CGF of energy exchange following Eq.~(\ref{eq:CF-TTM}) by performing simple algebraic manipulations of the Pauli matrices which yield \cite{Bijay-expt}
\begin{eqnarray}
 {\cal G}^{\rm spin}_{T}(u) &&= \ln \Big[1 + \sin^2 \Big(J T\Big) \Big\{f_1(\omega_0) \, (1 - f_2(\omega_0)) \big(e^{i u \omega_0} -1\big) \nonumber \\
&&+ f_2 (\omega_0)(1 - f_1(\omega_0)) \big(e^{-i u \omega_0} -1\big)\Big\}\Big],
\label{eq:chi-spin}
\end{eqnarray}
where $f_{i}(\omega_0)= (e^{\beta_i  \omega_0} + 1)^{-1}$, $i=1,2$ is the Fermi like distribution function. Once again the XFT is obeyed for arbitary $J, T, \beta_1$ and $\beta_2$ due to the thermal coupling symmetry. At this point, it is important to compare the CGF in eq.~(\ref{eq:chi-spin}) with the CGF  for the coupled oscillator (Eq.~(\ref{CGF:osc})). First of all, for both these models, interestingly the  $J\,T$ dependence appears in the same functional form ${\cal T}_{T}(J) =\sin^2 \big(J T\big)$. In fact, in this context it is simply the transition probability between states $|0 1 \rangle$ and $|1 0 \rangle$ i.e.,${\cal T}_{T}(J)= |\langle 1 0 | {\cal U}(T,0)| 0 1 \rangle |^2$ $\big($ $|0\rangle$ ($|1\rangle$) refers to the ground (excited) state for the qubit$\big)$. Second and most importantly, there are crucial sign differences in terms of the Bose and Fermi like functions, reflecting the key difference between a two-level spin system and an infinite-level harmonic oscillator system. 
In fact, because of this crucial sign change for the qubit setup, a looser bound for TUR appears, as we show below.  We once again write down the first two cumulants following the the CGF as
\begin{widetext}
\begin{align}
&\langle Q \rangle^{\rm spin} = \omega_0  {\cal T}_{T}(J) \Big[f_1 \!-\!f_2\Big], 
\\
&\langle  Q^2 \rangle ^{\rm spin}_{c}= \omega_0^2  \Big[{\cal T}_{T}(J) \Big( f_1 (1\!-\!f_2) \!+\! f_2 (1\!-\!f_1)\Big)\!-\! {\cal T}_{T}^2(J) \,\big(f_1\!-\!f_2\big)^2 \Big],\,\,\,\,\,\,  
\label{eq:analM}
\end{align}
\end{widetext}
Interestingly, for Fermi like function also a relation similar to Eq.~(\ref{equality-Bose}) exists, given as
\be
f_1 (1\!-\!f_2) \!+\! f_2 (1\!-\!f_1) = \coth \frac{\Delta \beta \omega_0}{2} (f_1 -f_2). 
\ee
This helps us to organize the cumulants and to receive the TUR ratio as
\be
\Delta \beta \frac{\langle  Q^2  \rangle^{\rm spin}_{c}}{\langle Q \rangle^{\rm spin}} = \Delta \beta \omega_0 \coth\big[\frac{\Delta \beta \omega_0}{2}\big] - \langle \Sigma \rangle^{\rm spin}.
\label{TUR-ratio-spin}
\ee
Once again this expression should be compared with Eq.~(\ref{TUR-ratio-HO}). The first term is the same as before. However, the apparent sign differences between the two models reflects in the second term  where the average entropy production term  appears as a negative contribution to TUR ratio. It is therefore not immediately obvious that,  this coupled two-qubit model will satisfy the T-TUR bound. In what follows we therefore first get an upper bound on the average entropy production and thereby provide a lower bound for the TUR.  Interestingly, this helps us to find a condition on  ${\cal T}_{T}(J)$ for which the T-TUR is respected.

%---------------figure------------------------
\begin{figure*}
\centering
\includegraphics[scale=1]{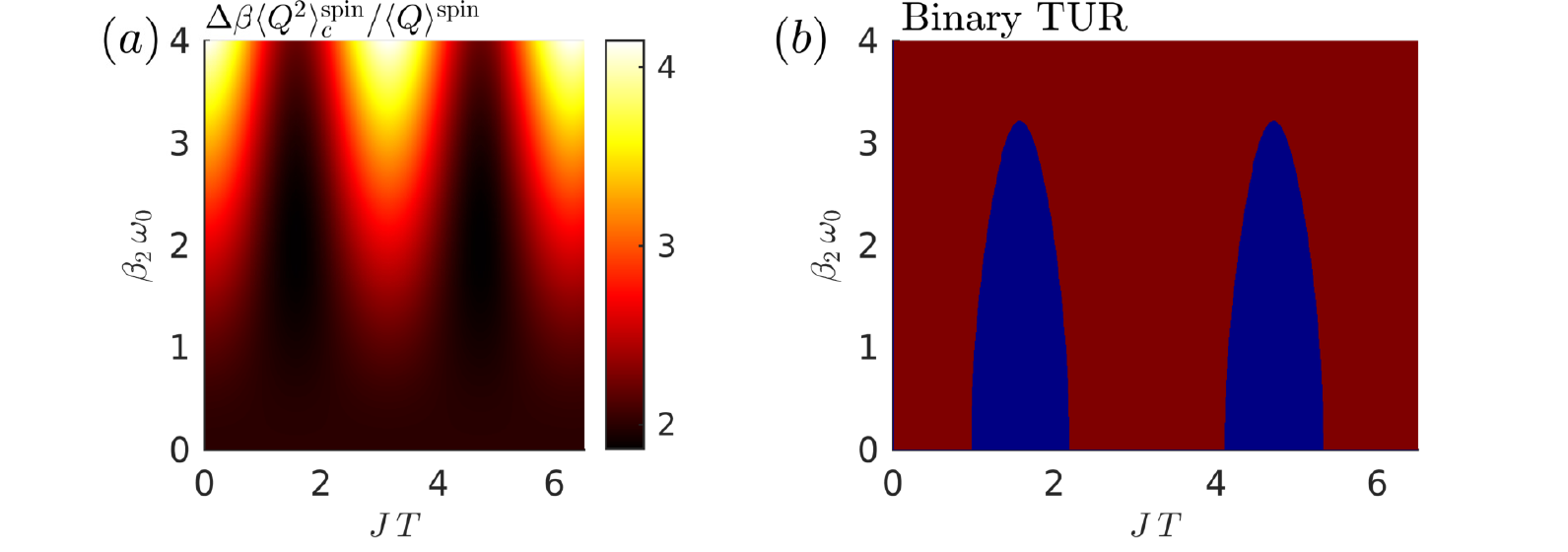}
\caption{(Color online): (a) Two-dimensional plot for TUR ratio ($\Delta \beta \frac{\langle  Q^2  \rangle^{\rm spin}_{c}}{\langle Q \rangle^{\rm spin}}$) for the coupled two-qubit system  as a function of $J \,T$ and $\beta_2 \omega_0$. We set $\beta_1=0$.  
(b) Corresponding binary plot of TUR. The violation (validity) regime of the T-TUR bound is colored by blue (dark red).}
\label{qubit_TUR_bin1}
\end{figure*}
%---------------figure------------------------

We first note that, the Fermi-like function can be alternatively written as, 
\be
f_i =\frac{1}{e^{\beta_i \omega_0} +1} = \frac{1}{2} \Big( 1- \tanh \frac{\beta_1 \omega_0}{2}\Big).
\ee
With the help of this expression, we write down the net entropy production as,
\bea
&&\langle \Sigma \rangle^{\rm spin}= \frac{\Delta \beta \omega_0 {\cal T}_{T}(J)}{2} \Big[\tanh \frac{\beta_2 \omega_0}{2}-\tanh \frac{\beta_1 \omega_0}{2}\Big] \nonumber \\
&&= \frac{\Delta \beta \omega_0 {\cal T}_{T}(J)}{2} \Big[\big(\tanh \frac{\Delta \beta \omega_0}{2}\big) \big(1\!-\!\tanh \frac{\beta_1 \omega_0}{2}\, \tanh \frac{\beta_2 \omega_0}{2}\Big]. \nonumber \\
\eea
Now since $\beta_i$ is always positive and $\tanh x$ is bounded function between (0,1) for $x>0$, the second term in the product in the above equation is always $<1$, which gives us
\be
\tanh \frac{\Delta \beta \omega_0}{2} \geq \tanh \frac{\beta_2 \omega_0}{2} - \tanh \frac{\beta_1 \omega_0}{2} 
\ee
and therefore, we receive an upper bound for the average entropy production,
\be
\langle \Sigma \rangle^{\rm spin} \leq \frac{\Delta \beta \omega_0 {\cal T}_{T}(J) }{2}\, \tanh \frac{\Delta \beta \omega_0}{2},
\ee
which finally translates to a lower bound on TUR ratio for this model as  
\be
\Delta \beta \frac{\langle  Q^2  \rangle^{\rm spin}_{c}}{\langle Q \rangle^{\rm spin}}  \geq  \Delta \beta \omega_0 \Big[\coth \frac{\Delta \beta \omega_0}{2} - \frac{{\cal T}_{T}(J)}{2} \tanh \frac{\Delta \beta \omega_0}{2} \Big].
\label{bound-qubit}
\ee
The equality sign here holds for $\beta_1=0$ or at equilibrium. Since the TUR ratio is periodic as a function of $J T$, we focus our attention within the first period $[0, \pi]$. The obtained bound indicates that, in the weak-coupling limit, i.e., $J\,T \ll 1$ which imply ${\cal T}_{T}(J) \ll 1$, the second term in the above expression can be ignored  and the T-TUR will be satisfied. In fact, it is easy to check that the T-TUR will remain to be valid for ${\cal T}_{T}(J) < 2/3$ which gives an allowed range for $J\,T$, ($ J T \leq 0.95$ and $J T \geq 2.19$, within the first period). Therefore, to observe a violation for the T-TUR, a necessary condition is to tune the value of $J\,T$ such that ${\cal T}_{T}(J) > 2/3$. However, note that, this condition is not a sufficient one. This can be seen as follows:  following the RHS of Eq.~(\ref{bound-qubit}), the minimum value for the TUR bound corresponds to ${\cal T}_{T}(J) =1$. Now for large $\Delta \beta$ ($\Delta \beta \, \omega_0 \gg 1)$, both $\coth$ and $\tanh$ functions saturate to value unity ($\Delta \beta \, \omega_0 \approx 6)$ which imply that the TUR bound scales as $\Delta \beta \omega_0/2$ and the T-TUR will be satisfied. Therefore, along with the condition ${\cal T}_{T}(J) > 2/3$, the violation of T-TUR in this case requires a careful tuning of $\Delta \beta \omega_0$. 

In Fig.~(\ref{qubit_TUR_bin1}(a)) we display a two-dimensional plot for TUR as a function of $\beta_2 \omega_0$ and $J\,T$. Fig~(\ref{qubit_TUR_bin1}(b)) is the corresponding binary plot differentiating the validity (dark-red) and the violation regimes (blue) of the T-TUR. We clearly observe a regime for which T-TUR is not valid and the results nicely matches with our theoretical predictions. As mentioned earlier, for sufficiently large $\Delta \beta$ ($\Delta \beta \omega_0 > 3.2$), the T-TUR bound is always satisfied. 
In contrast, the minimum value of the TUR bound is found to be $\approx 1.86$ which occurs for maximum transition probability ${\cal T}_{T}(J) = 1$ i.e,  $J\, T =\pi/2$ and $\Delta \beta \omega_0 \approx 2.01$.

%---------------figure------------------------
\begin{figure}
\centering
\includegraphics[trim= 0cm 0cm 0cm 0cm, clip=true,width=8cm]{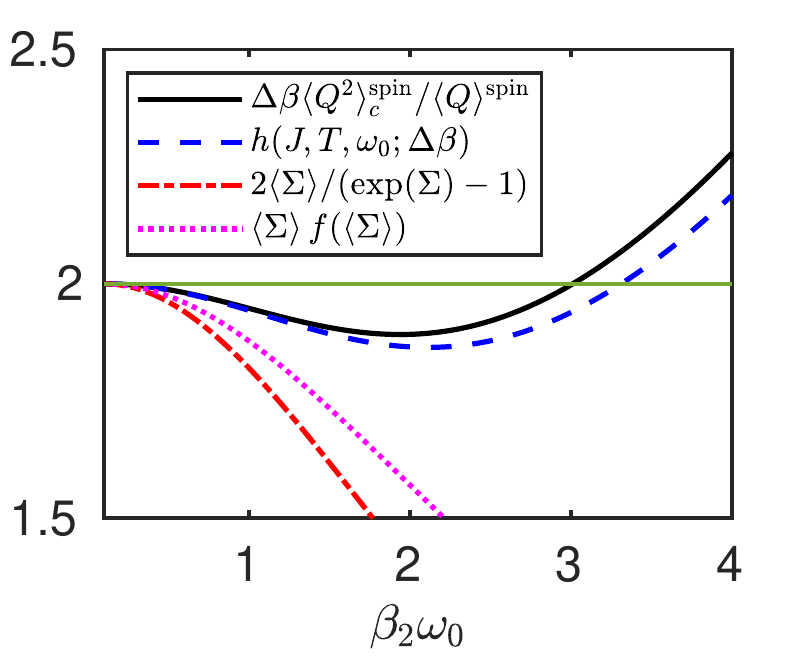} 
\caption{(Color online): Comparison between the TUR bounds obtained in Eq.~(\ref{bound-qubit}),(blue, dashed), denoted here by $h(J, T, \omega_0; \Delta \beta) = \Delta \beta \omega_0 \Big[\coth \frac{\Delta \beta \omega_0}{2} \!-\! \frac{{\cal T}_{T}(J)}{2} \tanh \frac{\Delta \beta \omega_0}{2}\big] $, the generalized bounds, GTUR-1 (red, dashed-dotted) in Eq.~(\ref{eq:fund-TUR-1}) and GTUR-2 (magenta, dotted) in Eq.~(\ref{eq:fund-TUR-2}) with the actual TUR value (black, solid). For reference a line is drawn at the value 2. The parameters are $\beta_1 \omega_0=0.1$, and $J\,T= \pi/2$. The bound in Eq.~(\ref{bound-qubit}) closely follow the actual TUR trend.}
\label{qubit_TUR_bound}
\end{figure}
%---------------figure------------------------

In Fig.~(\ref{qubit_TUR_bound}) we show that the TUR bound obtained in Eq.~(\ref{bound-qubit}) is in fact a tighter one compared to the generalized bound (Eq.~(\ref{eq:fund-TUR-1}) and Eq.~(\ref{eq:fund-TUR-2})). More importantly, we observe that the generalized bound obtained from fluctuation symmetry becomes loose with increasing $\Delta \beta$ whereas the obtained bound closely follow the actual TUR trend. In fact, for large $\Delta \beta$, the net entropy production $\langle \Sigma\rangle$ scales as $\Delta \beta$, hence the G-TUR1 behaves as $2 \langle \Sigma \rangle  /e^{\langle \Sigma \rangle }$ which tends to zero whereas the TUR bound obtained in Eq.~(\ref{bound-qubit}) scales as $\Delta \beta \omega_0 /2$. As expected, G-TUR2 performs a bit better than G-TUR1.

%---------------Subsection-Hybrid------------------------
\subsection{Hybrid spin-oscillator system}
As a final toy model example we consider a hybrid system consisting of a single qubit and a single quantum harmonic oscillator (see Fig.~(\ref{schematic}(c)), once again interacting via a thermal coupling term. The total Hamiltonian is given as 
\bea
{H}_{\rm JC} &=&  \frac{\omega_0}{2} \sigma_z \otimes 1_1  + 1_2 \otimes  {\omega_0} \, a^{\dagger} a 
\nonumber\\
&+& J\,(a^{\dagger} \otimes \sigma^{-} +  a \otimes  \sigma^{+}).
\label{eq:JC}
\eea 
where $\sigma^{\pm}= \sigma_x \pm i \sigma_y$ are the spin ladder operators.  
This model is in fact the famous Jaynes-Cummings (JC) model and is one of the most well studied setup in quantum optics.  We are interested here to analyze the quantum thermodynamics properties for this model and compute the exact CGF for the energy exchange. We provide here a brief outline of the derivation. 

Starting from Eq.~(\ref{eq:CF-TTM}) we switch to the interaction picture with respect to the bare part of the Hamiltonian (the first two terms of Eq.~(\ref{eq:JC})) and 
%------------------figure-----------------------
\begin{figure*}
\centering
\includegraphics[scale=1]{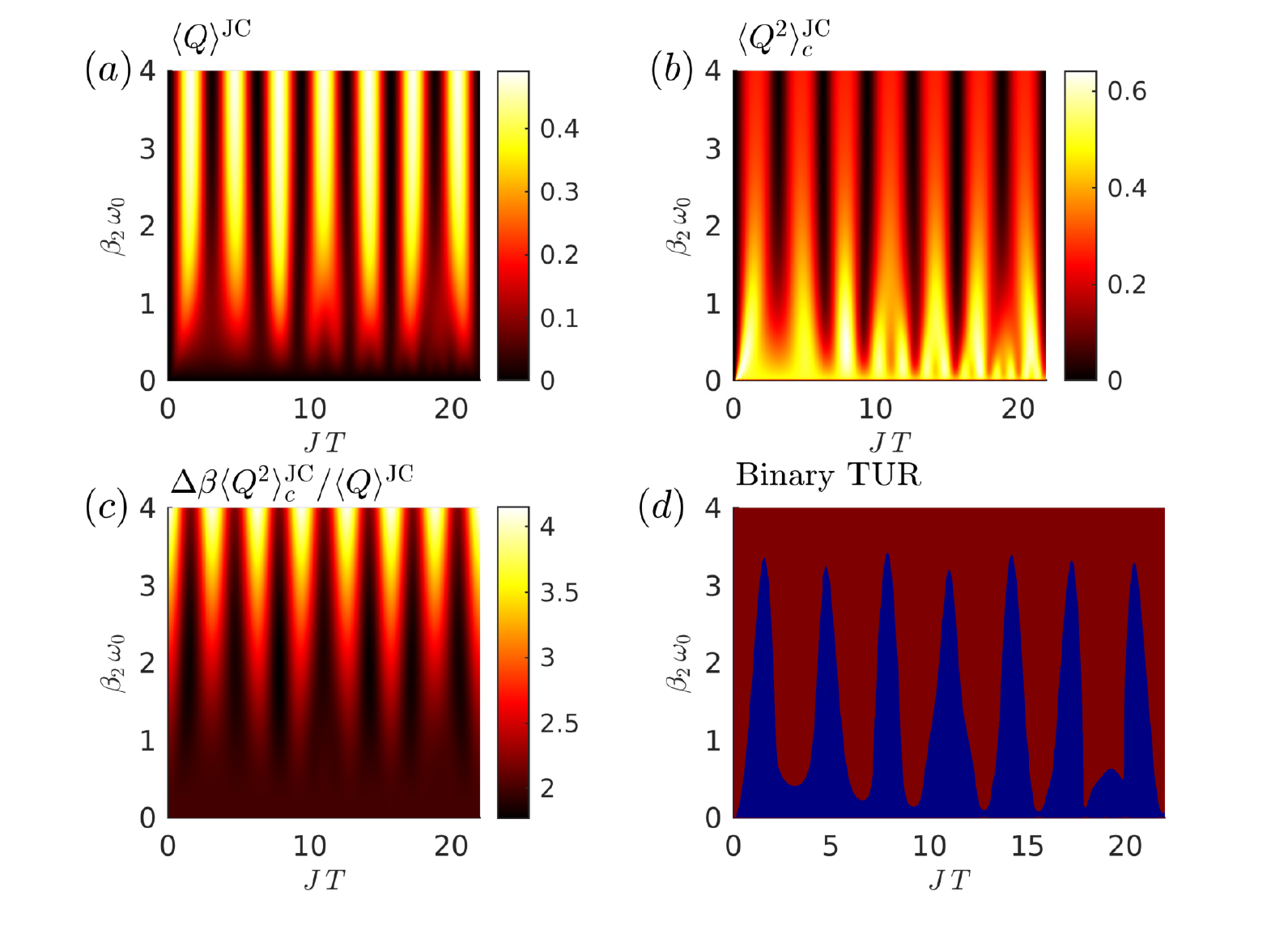} 
\caption{(Color online): Two-dimensional plots for the JC model: (a) average energy change $\langle Q \rangle^{\rm JC}$, (b) corresponding noise $\langle Q^2 \rangle^{\rm JC}_c$, (c) the TUR ratio $\Delta \beta \frac{\langle  Q^2  \rangle^{\rm JC}_{c}}{\langle Q \rangle^{\rm JC}}$ and (d) binary plot  of the same  TUR data where the violation (validity) regime of the T-TUR bound is colored by blue (dark red), as a function of $J T$ and $\beta_2 \omega_0$. We set $\beta_1=0$.} 
\label{JC_TUR_contour}
\end{figure*}
%------------------figure-----------------------
compute the total unitary propagator in the qubit basis. We receive \cite{Gerry}
\begin{equation}
U_{I}(t) = e^{- i V t}=
\begin{bmatrix}
\cos(\sqrt{a a^{\dagger}} J t) & - i \frac{\sin\big(\sqrt{a a^{\dagger}} J t\big)}{ \sqrt{ a a^{\dagger}}} a \\
- i \frac{\sin\big(\sqrt{a^{\dagger} a} J t\big)}{ \sqrt{ a^{\dagger} a }} a^{\dagger} & \cos(\sqrt{a^{\dagger}a} \, J t) 
\end{bmatrix},
\end{equation}
where we have used the convention that $\Big[U_{I}(t)\Big]_{11} = \langle e| U_{I}(t) | e \rangle$, $\Big[U_{I}(t)\Big]_{12} = \langle e| U_{I}(t) | g \rangle$, $\Big[U_{I}(t)\Big]_{21} = \langle g| U_{I}(t) | e \rangle$, and $\Big[U_{I}(t)\Big]_{22}= \langle g| U_{I}(t) | g \rangle$. 
Note that because of the commutable coupling condition, in the interaction picture, the time-ordered operator in the unitary propagator does not play any role. With the help of this exact unitary operator and carrying out the calculation in the qubit basis, the exact CGF can be written down as,
\begin{align}
{\cal G}^{\rm JC}_{T}(u)\!=\ln\Big[1 \! &+ {\cal Q} \,\, \Big\{\frac{f_1}{n_2} \big(e^{i u \omega_0} \!-\! 1\big)  + \frac{(1 - f_1)}{(1+n_2)} \big(e^{-i u \omega_0}\! -\!1 \big)\Big\}\Big],
\end{align}
where we define the function ${\cal Q}={\cal Q}(J, T, \omega_0; \beta_2)$ as
\be 
{\cal Q}(J, T, \omega_0; \beta_2)= \sum_{n=0}^{\infty} e^{-\beta_2 n \omega_0}\sin^2(\sqrt{n} J T). \label{eq: sum_QQ}
\ee
We make the following observations here: (i) Unlike the coupled oscillator or the coupled qubit model, for this hybrid setup the transition probability is weighted by the oscillator temperature $\beta_2$ as captured by the ${\cal Q}$ function.
(ii) Because of the hybrid nature of the setup both the Fermi like and the Bose functions appears in the CGF expression. 
Once again, it is easy to check the validity of XFT for arbitrary $J,T$ and the initial temperatures $\beta_1,\beta_2$. Note that, in the low-temperature limit of the oscillator i.e., $\beta_2 \omega_0 \gg 1$, it is expected that the above result should reproduce the two-qubit CGF. This can be seen as follows: for $\beta_2 \omega_0 \gg 1$ only $n=1$ term contributes to Eq.~(\ref{eq: sum_QQ}). Therefore, the ${\cal Q}$ function simplifies to  ${\cal Q} \approx e^{-\beta_2 \omega_0} \sin^2(Jt)$ and correspondingly the Bose functions simplifies to to $n_2 \approx e^{-\beta_2 \omega_0} \big(1+e^{-\beta_2 \omega_0}\big)$  and $1+n_2 \approx 1 + e^{-\beta_2 \omega_0}$ which gives ${\cal Q}/n_2 = (1-f_2) \sin^2 (J T)$ and ${\cal Q}/(1+n_2) = f_2 \sin^2 (J T)$ and thus one correctly recovers the two-qubit model CGF, given in Eq.~(\ref{eq:chi-spin}) 

We now investigate the TUR bound and write down the cumulants as
\begin{widetext}
\begin{align}
&\langle Q \rangle^{\rm JC} =\omega_0 \, {\cal Q} \, \Big(\frac{f_1}{n_2} - \frac{1-f_1}{1+n_2}\Big) = \omega_0 \, {\cal Q} \, \frac{f_1}{1+n_2} \Big(e^{\beta_2 \omega_0} - e^{\beta_1 \omega_0} \Big),
\label{current:JC}
\\
&\langle  Q^2  \rangle^{\rm JC}_{c}= \omega_0^2 \, {\cal Q}\, \Big[\Big(\frac{f_1}{n_2} + \frac{1-f_1}{1+n_2}\Big)- {\cal Q} \Big(\frac{f_1}{n_2} - \frac{1-f_1}{1+n_2}\Big)^2 \Big].
\label{noise:JC}
\end{align}
\end{widetext}
As expected, the energy exchange in Eq.~(\ref{current:JC}) vanishes when both the spin and the oscillator are initially kept at the same temperature. Interestingly, we once again receive a similar identity as in Eq~(\ref{bose-relation}) but now involving both the Fermi and Bose functions,
\be
f_1 ( 1+ n_2) \! + n_2 (1-f_1) = \coth\Big[ \frac{\Delta \beta \omega_0}{2}\Big]  \big(f_1 ( 1+ n_2) \!-\! n_2 (1-f_1)\big),
\ee 
which helps us to write the TUR ratio as 
\be
\Delta \beta \frac{ \langle Q^2  \rangle^{\rm JC}_{c}}{\langle Q \rangle^{\rm JC}} = \Delta \beta \omega_0 \coth \frac{\Delta \beta \omega_0}{2} - \langle \Sigma \rangle^{\rm JC}.
\ee
This expression once again should be compared with Eq.~(\ref{TUR-ratio-HO}) and Eq.~(\ref{TUR-ratio-spin}).  Interestingly, analogous to the previous cases, the first term remains the same. Whereas, the average entropy production term for the hybrid case produces a negative contribution to the TUR ratio, as was the case for the two-qubit model.  Therefore, the breakdown of the T-TUR bound can be expected even for this setup. However, note  that, in the limit when ${\cal Q} \ll 1$, i.e., in the weak-coupling limit, the T-TUR is once again preserved. In Appendix B we provide a general proof for the bound for any two weakly coupled system.

In Fig.~(\ref{JC_TUR_contour}) we display the two-dimensional plots for the first and second cumulant and the corresponding TUR ratio. Notice that, the cumulants and the corresponding TUR ratio in not entirely periodic as a function of $J\,T$, especially in the high-temperature regime $\beta_2 \omega_0 \ll 1$. This is clear from the expression for the function ${\cal Q}$.The violation for the T-TUR bound is clearly observed in the binary plot (Fig.~(\ref{JC_TUR_contour}(d)).
Expectedly, the low temperature behaviour for the TUR ratio is found to be similar with the two-qubit case with clear validity of T-TUR bound beyond $\Delta \beta \omega_0 \approx 3.4$.  However, in the high temperature regime the violation regime for the JC model is broader (comparing TUR ratio vs $J\,T$ within the first period in both Fig.~(\ref{qubit_TUR_bin1}(b)) and Fig.~(\ref{JC_TUR_contour}(d)) ) in comparsion to the two-qubit case. This is because of the availability of many states for the oscillator leading a significant contribution of the average entropy production.

%------------------Section-----------------------
\section{Summary.}
%---------------------------------------------------
We examined the TUR bound for energy exchange for three simple model systems characterized by different underlying statistics for the transport carriers. We obtained exact analytical expressions  for the heat exchange characteristic function for all three cases which hands over the cumulants to analyze the TUR. One of the interesting observations was the similarity in the expressions for the CGF for the two-qubit and two-oscillator model where they differ by crucial sign differences arising from the underlying Fermi-like and the Bose statistics. We found that, in general, the TUR ratio is sensitive to the statistics and the validity/violation of the T-TUR  is critically dependent on this. In all three cases, interestingly, the TUR ratio was organized in terms of an universal term which is always greater or equal to 2 and a net entropy production term. The deviation from the T-TUR bound largely depends on the contribution of this average entropy production to the TUR ratio. For coupled oscillator system, displaying pure bosonic statistics, this contribution turned out to be always positive and thus the tighter bound is always preserved. In contrast, the appearance of the Fermi-like statistics for both the qubit and the the hybrid model leads to a negative contribution, leading to a lower bound (smaller than the T-TUR) for the TUR. 
%The violation in more prominent for the JC model due to the possibility of significant entropy production 
However, in the weak-coupling regime, all these models satisfy the T-TUR bound. Future work will direct towards designing finite time heat engine cycles based on these toy models and understand the impact of the statistics on the engine efficiency and the corresponding TUR bound.  

%In other words, there is a threshold value of the coupling beyond which violation of the T-TUR can be observed. 

%We showed that for As a first key result, we showed that,  in the weak coupling limit between the systems, the T-TUR always satisfied. In other words, there is a threshold value of the coupling beyond which S-TUR violation can be observed but that depends on the statistics. 

%The S-TUR thus contains practical information: The condition to invalidate it pinpoints to
%regimes of favorable performance for heat machines, operating with high constancy {\it and} little dissipation.

%As expected, the generalized version of the TUR, originating from the underlying fluctuation symmetry is respected in all these setups 

\section {\bf Acknowledgments}
BKA gratefully acknowledges the financial support from Max Planck-India mobility grant. BKA thanks D. Segal for useful discussions. SS acknowledge support from the Council of Scientific \& Industrial Research (CSIR), India (Grant Number 1061651988). OS acknowledges  the Inspire grant from Department of Science \& Technology (DST), India.

%-----------------APPENDIX--------------------------

\renewcommand{\theequation}{A\arabic{equation}}
\setcounter{equation}{0}  % reset counter

\section*{Appendix A: Exchange fluctuation theorem (XFT) under commutable coupling condition}
In this Appedix we prove that for a bipartite setup under the thermal coupling limit the XFT is valid for arbitrary coupling strength between two systems and for arbitrary time duration of energy exchange. The starting point here is the CF for energy exchange, given in  Eq.~(\ref{eq:CF-TTM}),
\begin{eqnarray}
\!\!\chi_{T}(u)&\!=\!{\rm Tr}\Big[{\cal U}^{\dagger}(T,0) (e^{-i u H_1} \otimes {1}_2)\, {\cal U}(T,0)  (e^{i u H_1} \otimes {1}_2) \rho(0)\Big]\nonumber,\\
\label{eq:CF-TTM1}
\end{eqnarray}
where ${\cal U}(t,0) = e^{ - i H t}$ is the global unitary operator with $H=H_1 + H_2 +V$. To take advantage of the thermal coupling limit i.e., the commutable coupling condition $[H_1+ H_2, V]=0$, one can rewrite the above expression along with the consideration that initially both the systems are in their respective Gibbs thermal state, i.e.,  $\rho(0)= \frac{e^{-\beta_1 H_1}}{{\cal Z}_1} \otimes \frac{e^{-\beta_2 H_2}}{{\cal Z}_2} $ which further imply $[\rho(0), H_1]=0$. One then receives,
\bea
\!\!\!\!\chi_{T}(u)\!\!=\!\ \!   && \frac{1}{{\cal Z}_1 {\cal Z}_2} {\rm Tr}\Big[(e^{i u H_1} \otimes {1}_2)\, e^{\Delta \beta H_1}\, e^{-\beta_2 ( H_1+H_2)} \, {\cal U}^{\dagger}(T,0) \nonumber \\
&&  \, \, \quad \quad   (e^{-i u H_1} \otimes {1}_2)\, {\cal U}(T,0)\Big],
\eea
where recall that $\Delta \beta= \beta_2-\beta_1$.  The thermal coupling condition allows to swap the third and  the fourth term. Next performing cyclic permutation under the trace operation, we receive,
\bea
\chi_{T}(u) &=& \frac{1}{{\cal Z}_1 {\cal Z}_2} {\rm Tr}\Big[{\cal U}(T,0) (e^{-i(-u+i \Delta \beta) H_1} \otimes {1}_2)\, \nonumber \\
&&{\cal U}^{\dagger}(T,0) (e^{i(-u+i \Delta \beta) H_1} \otimes {1}_2)\, \rho(0)\Big],
\eea
where  $\Delta \beta= \beta_2-\beta_1$. This expression still does not give us the XFT that we are looking for. In fact, at this point the above expression satisfy a fluctuation relation given as $\chi_{T}(u)= \chi_{-T}(-u + i \Delta \beta)$ connecting forward and reversed protocol.

In order to proceed, we now assume that the the composite and the individual systems are time reversal invariant, which is the case considered in this paper. We then have $\Theta \,{\cal U}(T,0) ={\cal U}^{\dagger}(T,0)\, \Theta $,
$\Theta \, e^{-i u H_1} = \,e^{i u^* H_1} \Theta$ and $[\Theta\,,\,H_{1,2}]=0$.  
$\Theta$ is time reversal operator. Now inserting $\Theta^{-1} \Theta$ inside the trace and using eq. (A2) we receive
\bea
\chi_{T}(u)&=&\frac{1}{{\cal Z}_1 {\cal Z}_2} {\rm Tr}\Big[\Theta^{-1} \Theta{\cal U}(T,0) 
(e^{-i(-u+i \Delta \beta) H_1} \otimes {1}_2)\,\nonumber \\ 
&& \quad \quad {\cal U}^{\dagger}(T,0) (e^{i(-u+i \Delta \beta) H_1} \otimes {1}_2)\, \rho(0)\Big],\nonumber \\ 
&& =\frac{1}{{\cal Z}_1 {\cal Z}_2} {\rm Tr}\Big[\Theta^{-1} {\cal U}^{\dagger}(T,0)
(e^{i(-u^*-i \Delta \beta) H_1} \otimes {1}_2)\, \nonumber \\
&& \quad \quad {\cal U}(T,0) (e^{-i(-u^*-i \Delta \beta) H_1} \otimes {1}_2)\, \rho(0) \Theta \Big].
\eea
Now due to the antilinear nature of $\Theta$ we have ${\rm Tr}[\Theta^{-1} \,A\, \Theta] ={\rm Tr}[A^{\dagger}] \cite{fluct2} $. Therfore we finally receive
\bea
\chi_{T}(u) =\frac{1}{{\cal Z}_1 {\cal Z}_2} {\rm Tr}\Big[{\cal U}^{\dagger}(T,0) (e^{-i(-u+i \Delta \beta) H_1} \otimes {1}_2)\, \nonumber \\
 {\cal U}(T,0) (e^{i(-u+i \Delta \beta) H_1} \otimes {1}_2)\, \rho(0)\Big]  \nonumber \\
 = \chi_{T}(-u+i\Delta \beta) \quad 
\eea
for arbitrary time duration $T$ and coupling strength.

\renewcommand{\theequation}{B\arabic{equation}}
\setcounter{equation}{0}
\section*{Appendix B: Proof of T-TUR in the weak coupling regime}
\begin{center}
\begin{figure}
\includegraphics[trim=8cm 10cm 10cm 3cm, clip=true,width=\columnwidth]{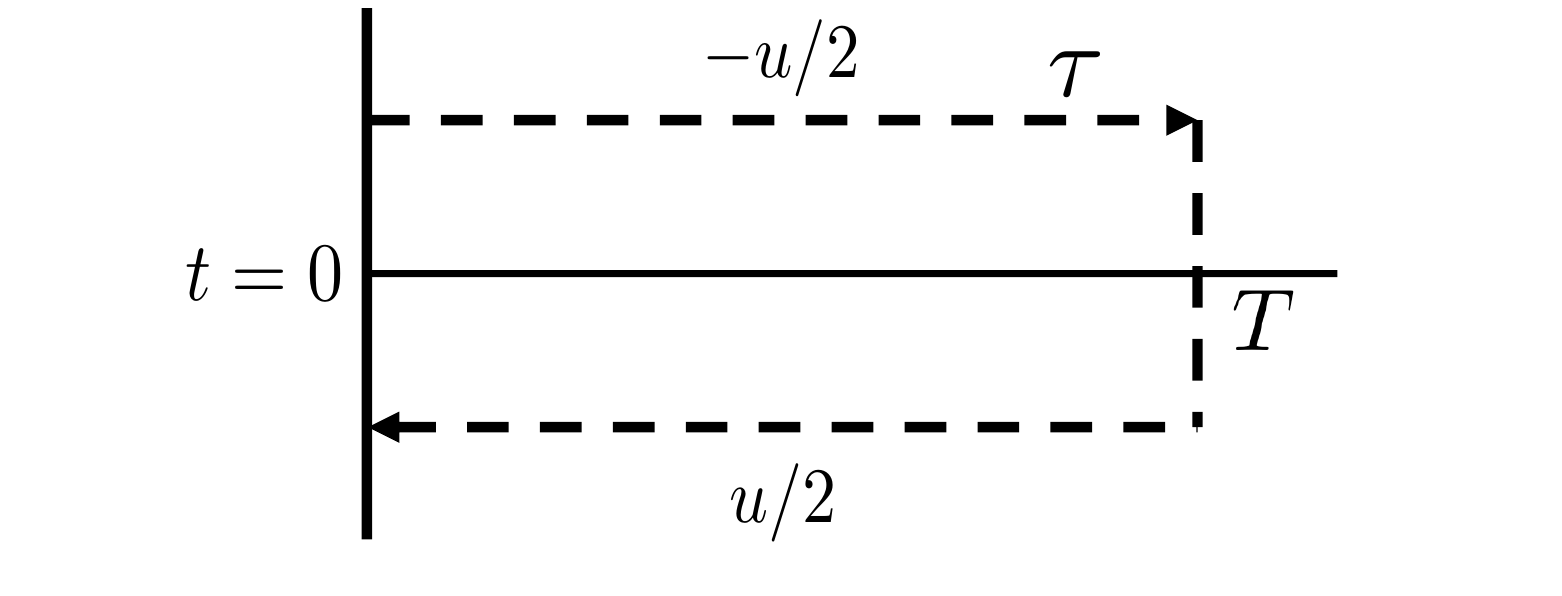}
\caption{The complex time Keldysh Contour with upper and lower branch. The contour path begins from $t=0$, goes to maximum time $t=T$ and then comes back to $t=0$ again. The upper (lower) branch corresponds to time-ordered forward  (anti-time-ordered backward) evolution propagator.  For energy counting statistics problem the Hamiltonian is dressed differently in the upper ($-u/2$) and the lower ($+u/2$) branch by the counting parameter $u$. $\tau$ is the contour-time parameter.}
\label{contour}
\end{figure}
\end{center}

In this Appendix we provide a proof for the tighter bound of TUR (T-TUR) in the weak coupling regime for generic bipartite setup. We employ here the Keldysh non-equilibrium Green's function approach \cite{NEGF1, NEGF2, NEGF3} for the proof. This method is useful to receive a bound in transient as well as in the steady-state regime,  as we show below. We begin with Eq.~(\ref{eq:CF-TTM}) and organize the characteristic function in the interaction picture as,
\begin{eqnarray}
\!\!\!\!\!\!\chi_{T}(u)&\!\!=\!\!&\int dQ \, e^{i u Q} \, p_T(Q), \nonumber \\
=&&\!\!{\rm Tr}\Big[{\cal U}_{I} ^{\dagger}(T,0) (e^{-i u H_1} \otimes {1}_2) {\cal U}_{I} (T,0)  (e^{i u H_1} \otimes {1}_2) \rho(0)\Big],
\nonumber 
\label{eq:CF-TTM-app}
\end{eqnarray}
where $U_I(t,0)= {\cal T} \exp \big[-i \int_{0}^{t} V_{I} (t') dt'\big]$ with ${\cal T}$ being the time-ordered operator and $V_{I}(t) = e^{i H_0 t} \, V \, e^{-i H_0 t}$, $H_0=H_1+H_2$.  Recall that, the composite density matrix is decoupled at the initial time $t=0$ with each system is in thermal equilibrium at a particular temperature i.e., $\rho(0)=\rho_1 \otimes \rho_2 =\frac{e^{-\beta_1 H_1}}{{\cal Z}_1} \otimes \frac{e^{-\beta_2 H_2}}{{\cal Z}_2} $. This condition imply $\big[\rho(0), H_0] =0$. The above equation then can be organized as
\begin{equation}
\chi_{T}(u)={\rm Tr}\Big[ \rho(0) \,\big[U_{I}^{\dagger}\big]^{u/2}(T,0) \, U_{I}^{-u/2}(T,0) \Big],
\end{equation}
where now both the forward and backward evolution operators are dressed by the counting field $u$. This expression can  be recast on a Keldysh contour as (see Fig.~(\ref{contour}))
\be
\chi_{T}(u) \,=\, {\rm Tr} \Big[ \rho(0) T_c e^{-{i} \int_c V_{I}^x(\tau) d\tau}  \Big],
\label{contour-eq}
\ee
where $T_c$ is the contour-ordered operator, which orders  operators according to their contour time argument: an earlier (later) contour time places the operator to the right (left). Therefore, the upper (lower) branch  corresponds to the forward (backward) evolution. $x(\tau)$ is a contour time dependent function which can take two possible values depending on the location of $\tau$ on the contour branch. Here $x^{+}(t) = - u/2$ for the upper branch (denoted by the $+$ sign) and  $x^{-}(t) = u/2$ for the lower branch (denoted by the $-$ sign) within the measurement time interval $[0,\tau]$. $x^{\pm}(t)=0$ outside the measurement time. Finally $V_I^x(\tau)= e^{i x H_1} V_I(\tau)  e^{-i x H_1}$ is the modified contour-time dependent operator dressed by the counting field. 

Often, instead of the CF, it is more convenient to work with the logarithm of the characteristic function ${\cal G}_{T}(u) \equiv \log \chi_{T}(u)$ which according to the linked-cluster theorem \cite{Mahan} contains only the connected diagrams. Since our focus is in weak-coupling regime, we therefore expand the exponential and collect terms up to the leading order in the coupling $V$ that produces non-zero contribution.  It turns out that the first order contribution in $V$ vanishes. This can be shown as follows:
The CGF in the first order, denoted by $ {\cal G}_{T} ^{(1)}(u) $  is given as
\bea
{\cal G}_{T} ^{(1)}(u) &=& -{i} \int d\tau  \langle V_{I}^x(\tau) \rangle \nonumber \\
&=&  -{i}\, \int_{0}^{T} dt \Big[\langle V_{I}^{x^{+}}(t) \rangle - \langle V_{I}^{x^{-}}(t) \rangle \Big],
\eea
where in the second line we transform back to the real time ($t$) from the contour time ($\tau$) using the Langreth's rule \cite{NEGF3,NEGF4}. Note that, in this order the contour-ordered operator does not play any role. 
%that says $\int d\tau f(\tau) = \int dt \big[f^{+}(t) -
Now since $V_{I}^{x^{\pm}}(t_1) = e^{ \mp i \xi/2 H_1} V_{I}(t)  e^{ \mp i \xi/2 H_1}$ and furthermore because $\big[\rho(0), H_1\big]=0$, the counting field dependent phase factors  cancels out exactly leaving $\langle V_I^{x^{+}}(t)\rangle =\langle V_I^{x^{-}}(t)\rangle$, i.e., independent of the branch index and thus the above contribution vanishes.

Next, the second order contribution to the CGF is given as,
\bea
{\cal G}_{T} ^{(2)}(u)  &=& \frac{(-i)^2}{2} \, \int d\tau_1 \int d\tau_2\,  \langle T_c\, V_{I}^x(\tau_1) \, V_{I}^x(\tau_2) \rangle_c \nonumber\\
&=& \int d\tau_1 \, \int d\tau_2 \, \tilde{G}_{c}(\tau_1,\tau_2),
\eea
where $\tilde{G}_{c}(\tau_1,\tau_2)$ indicates the connected part of the correlation function with the tilde symbol referring to the counting field dependence. Since the normalization condition demands that ${\cal G}^{(2)}_{T}(u\!=0)=0$, one can explicitly enforce the normalization in the above expression as
\be
{\cal G}_{T}^{(2)}(u) =  \int d\tau_1 \int d\tau_2 \, \Big[\tilde{G}_{c}(\tau_1,\tau_2)- {G}_{c}(\tau_1,\tau_2) \Big],
\label{central-proof}
\ee
where recall that Green functions without the tilde symbol refers to $u=0$. We once again transform back to the real time following the same procedure as mentioned earlier and obtain, 
\bea
{\cal G}_{T}^{(2)}(u)& =&  \int_{0}^{T} dt_1 \int_{0}^{T}  dt_2 \, \Big[{G}_{c}^{<}(t_1,t_2) + {G}_{c}^{>}(t_1,t_2) \nonumber \\
&& -\tilde{G}_{c}^{<}(t_1,t_2)-\tilde{G}_{c}^{>}(t_1,t_2)\Big],
\eea
where  $<$  ($>$) symbol corresponds to the lesser (greater) component of the Green function.  In order to proceed from here, we choose a generic form for the coupling, given as  $V= J \, A \otimes B  $, where $A \,(B)$ corresponds to Hermitian operator involving system 1 (system 2). For simplicity, we consider single degree of freedom and systems with Bosonic degree of freedom. However, the calculation can be straightforwardly extended for fermionic as well as for hybrid systems.
%For example, for the two coupled oscillator case, $A=a$, $B=b^{\dagger}$ and $A=a^{\dagger}$, $B=b$. $\lambda$ is the coupling constant.  The counting field dependence appears as $V_{I}^x(\tau) = J \,  \tilde{A}_I(\tau) \otimes B_I(\tau)$.
Now since the average in the Green functions are taken over $\rho(0)$ i.e., decoupled initial state,  the connected part of the correlation function in the contour-time reduces to 
\bea
\tilde{G}_{c}(\tau_1,\tau_2)= \frac{J^2}{2} \, \tilde{g}_A(\tau_1,\tau_2) \, g_B(\tau_2,\tau_1),
\eea
where $\tilde{g}_A(\tau_1,\tau_2) = - i \, \langle T_c A^{x}(\tau_1) A^{x}(\tau_2) \rangle$ is the bare but counting field dependent correlation function for system 1 with average taken over the equilibrium density operator $\rho_1=\frac{e^{-\beta_1 H_1}}{{\cal Z}_1} $ and similarly ${g}_B(\tau_1,\tau_2) = - i \, \langle T_c B (\tau_1) B(\tau_2) \rangle$ with average taken over the equilibrium density operator $\rho_2=\frac{e^{-\beta_2 H_2}}{{\cal Z}_2} $. 
Following Eq.~(\ref{central-proof}), in the real time we are interested only in the lesser and the greater components which are given as
\bea
\tilde{G}^{<}_c(t_1,t_2) &=& \frac{J^2}{2} \, g_A^{<}(t_1\!-\!t_2 - u) \, g_B^{>}(t_2-t_1), \nonumber \\
\tilde{G}^{>}_c(t_1,t_2) &=& \frac{J^2}{2} \, g_A^{>}(t_1\!-\!t_2 + u) \, g_B^{<}(t_2-t_1).
\eea
%Once again, recall that, the Green's functions without the tilde symbol corresponds to $u=0$. 
Since each of the bare Green functions are time-translational invariant, we can work in the frequency domain by performing Fourier transformation which gives,
\bea
&& \chi_{T}(u) = -\frac{J^2}{2} \int_{-\infty}^{\infty} \frac{d\omega_1}{2\pi}\, \int_{-\infty}^{\infty} \frac{d\omega_2}{2\pi} \,\, \frac{\sin^2 \big[\frac{(\omega_1-\omega_2)T}{2}\big]}{\frac{(\omega_1-\omega_2)^2}{4}} \nonumber \\
&& \,\,\,\,\, \Big[g_A^{<}(\omega_1) g_B^{>}(\omega_2) \big(e^{i u \omega_1} \!-\!1\big) \!+\!  g_A^{>}(\omega_1) g_B^{<}(\omega_2) \big(e^{-i u \omega_1} \!-\!1\big)\Big].\,\,\,\, \,\, \quad 
\label{final-CGF}
\eea
%We now consider two specific situations: 
%(i) To ensure, XXXXX 
Notice that since bare Green functions are computed with respect to their respective equilibrium state, they follow the standard Kubo-Martin-Schwinger boundary condition \cite {NEGF4} given as $g^{>}_{A}(\omega) = e^{\beta_1 \omega} g^{<}_A(\omega)$ and similarly for system 2 Green function  $g^{>}_{B}(\omega) = e^{\beta_2 \omega} g^{<}_B(\omega)$. Using this condition, one can rewrite the expressions for first and second cumulant by taking derivative of Eq.~(\ref{final-CGF}) with respect to $iu$ and receive,
\bea
\langle Q \rangle &=& - \frac{J^2}{2} \int_{-\infty}^{\infty} \frac{d\omega_1}{2\pi}\, \, \int_{-\infty}^{\infty} \frac{d\omega_2}{2\pi}\, \omega_1 \, {\cal F}(\omega_1,\omega_2; T) \nonumber \\ 
&&\, g_A^{>}(\omega_1) g_B^{<}(\omega_2) \Big[e^{-\beta_1 \omega_1}\, e^{\beta_2 \omega_2} -1\Big], \quad \quad  \\
\langle Q^2 \rangle_c &=& \! -\frac{J^2}{2}  \int_{-\infty}^{\infty} \frac{d\omega_1}{2\pi}\, \, \int_{-\infty}^{\infty}\, \frac{d\omega_2}{2\pi}\, \omega_1^2 \, {\cal F}(\omega_1,\omega_2; T) \nonumber \\ \, 
&& g_A^{>}(\omega_1) g_B^{<}(\omega_2) \Big[e^{-\beta_1 \omega_1}\, e^{\beta_2 \omega_2} +1\Big], \quad  \quad 
\eea
where we define ${\cal F}(\omega_1,\omega_2; T)= \frac{\sin^2 \big[\frac{(\omega_1-\omega_2)T}{2}\big]}{\frac{(\omega_1-\omega_2)^2}{4}}$. 
Up to this point the only approximation made was the weak-coupling. However, this does not automatically ensure current conservation or the XFT, as reflected in the above equation. In order to meet these criteria one needs to further impose resonant condition for energy exchange. In order to achieve this,  we use many-body quantum state representation for the individual system Hamiltonian $H_1$, and $H_2$, and write the lesser and greater components of the Green functions explicitly. For system 1,
\bea
g_{A}^{<}(t) & =& - i \sum_{m n} \frac{e^{-\beta_1 E_m}}{Z_A} |A_{m,n}|^2 e^{i \omega_{nm} t},\nonumber \\
g_{A}^{>}(t) & =& - i \sum_{m n} \frac{e^{-\beta_1 E_m}}{Z_A} |A_{m,n}|^2 e^{-i \omega_{nm} t},
\eea
where $\omega_{nm} = E_n-E_m$, $|A_{mn}|^2= |\langle m | A |n \rangle |^2$ with $|m\rangle,|n\rangle$ being the energy eigenstates for system 1 with Hamiltonian $H_1$ and $E_m$, $E_n$ are the corresponding eigenvalues. One receives similar expression for $g_{B}^{<,>}(t)$ but with inverse temperature $\beta_2$. We denote the corresponding energy eigenstates with $|p\rangle, |q\rangle$. Using Fourier transformed version of these Green functions we receive for the average energy change 
\bea
\langle Q \rangle &=& 2 \pi^2 J^2 \sum_{m n}\sum_{pq} \, \omega_{mn} \, {\cal F}(\omega_{mn},\omega_{qp}; T) |A_{mn}|^2\, |B_{pq}|^2\,  \nonumber \\
&& \Big[e^{-\beta_1 \omega_{mn}} \, e^{\beta_2 \omega_{qp}} - 1\Big] \frac{e^{-\beta_1 E_m}}{Z_A}\,\frac{e^{-\beta_2 E_p}}{Z_B}.
\eea
We now impose the resonant energy exchange condition between the two systems which imply $E_m-E_n \approx E_q -E_p$ i.e., $\omega_{mn} \approx \omega_{qp}$ leading to ${\cal F}(\omega_{mn},\omega_{qp}; T) \approx  T^2 $ and we receive,
\bea
\langle Q \rangle &=& 2 \pi^2 J^2  T^2 \sum_{m n}\sum_{pq} \omega_{mn} \, |A_{mn}|^2\, |B_{pq}|^2\,  \nonumber \\
&& \Big[e^{\Delta \beta \omega_{qp}} - 1\Big] \frac{e^{-\beta_1 E_m}}{Z_A}\,\frac{e^{-\beta_2 E_p}}{Z_B}.
\eea
Using the same resonant condition, we receive for the noise 
\bea
\langle Q^2 \rangle_c &=& 2 \pi^2 J^2  T^2 \sum_{m n}\sum_{pq} \omega_{mn}^2 \, |A_{mn}|^2\, |B_{pq}|^2\, \frac{e^{-\beta_1 E_m}}{Z_A}\,\frac{e^{-\beta_2 E_p}}{Z_B} \nonumber \\
&& \times \Big[e^{\Delta \beta \omega_{qp}} + 1\Big]  \\
&=&  2 \pi^2 J^2  T^2 \sum_{m n}\sum_{pq} \omega_{mn}  \, |A_{mn}|^2\, |B_{pq}|^2\,   \frac{e^{-\beta_1 E_m}}{Z_A}\,\frac{e^{-\beta_2 E_p}}{Z_B} \nonumber \\
&& \times\,\, \omega_{qp}\, \coth\big[ \frac{\Delta\beta \omega_{qp}}{2} \big] \,  \Big[e^{\Delta \beta \omega_{qp}} - 1\Big],\nonumber \\
&& \geq \frac{2}{\Delta \beta} \langle Q \rangle
\eea
where going from the first to the third line we write $\omega_{mn}^2 \approx \omega_{mn} \, \omega_{qp}$. Notice the important term $\omega_{qp} \,\coth\big[ \frac{\Delta\beta \omega_{qp}}{2} \big]$ in the fourth line which is always greater or equal to $2/\Delta \beta$ using which we receive the T-TUR bound. Also, note that the cumulants in this limit scales with $T^2$ and the entire analysis remains valid for $JT\ll 1$. 
%in the weak-coupling regime.  

The another key importance of the expression in Eq.~(\ref{final-CGF}) is that one can readily discuss results for the long-time limit $T\to \infty$. In fact, if a unique long-time limit of Eq.~(\ref{final-CGF})  exists that supports a non-equilibrium steady-state for the bipartite setup (imagining each system to be macroscopic bath) in which case all cumulants scale with $T$, as 
\be 
\lim_{T \to \infty} \,  \frac{\sin^2 \big[\frac{(\omega_1-\omega_2)T}{2}\big]}{\frac{(\omega_1-\omega_2)^2}{4}}  = 2 \pi T \delta(\omega_1-\omega_2)
\ee
and one receives the following expressions for the cumulants following Eq.~(\ref{final-CGF})
\bea
\frac{\langle Q \rangle}{T} &=& -J^2 \int_{-\infty}^{\infty} \frac{d\omega}{4\pi}\, \omega \,  g_A^{>}(\omega) g_B^{<}(\omega)\Big[ e^{\Delta \beta \omega} -1 \Big], \quad \quad  \quad \\
\frac{\langle Q^2 \rangle_c}{T} &=& - J^2 \int_{-\infty}^{\infty} \frac{d\omega}{4\pi}\, \omega^2 \, g_A^{>}(\omega) g_B^{<}(\omega)\Big[ e^{\Delta \beta \omega} + 1 \Big] \quad \quad  \quad \nonumber \\
&=& -J^2 \int_{-\infty}^{\infty} \frac{d\omega}{4\pi}\, \omega^2 \coth\Big[\Delta \beta \omega/2\Big] \, g_A^{>}(\omega) g_B^{<}(\omega) \nonumber \\
&& \times \Big[ e^{\Delta \beta \omega} - 1 \Big],\nonumber \\
&\geq& - \frac{2}{\Delta \beta}\,  J^2\, \int_{-\infty}^{\infty} \frac{d\omega}{4\pi}\, \omega  \, g_A^{>}(\omega) g_B^{<}(\omega)\Big[ e^{\Delta \beta \omega} - 1 \Big]\nonumber \\
&\geq& \frac{2}{\Delta \beta} \frac{\langle Q \rangle}{T}
\eea
where once again, like in the previous case, in the third line of $\langle Q^2\rangle_c /T$ expression we use the inequality $\omega \coth[ \Delta \beta \omega/2] \geq 2 / \Delta \beta$. Therefore for weakly coupled bipartite setup in the steady-state the T-TUR is preserved. It is crucial to note that both the G-TUR1 and G-TUR2 in the long-time limit fails to predict any non-trivial bound for the TUR ratio as the average entropy-production $\langle \Sigma \rangle$ diverges as $T\to \infty$.

\renewcommand{\theequation}{C\arabic{equation}}
\setcounter{equation}{0}
\section*{ Appendix C: Derivation of the exact CGF for the two-oscillator system}
In this Appendix we provide the derivation for the exact CGF given in Eq.~(\ref{CGF:osc}). We once again empoly the Keldysh non-equilibrium Green's function approach to derive the CGF. Note that this powerful approach can be extended to study bilinear systems with arbitrary complexity (See Ref.~(\onlinecite{Bijay12}) for details). 
%As before, the starting point is the CF given in Eq.~(\ref{eq:CF-TTM1}). In the interaction picture, with respect to the bare part of the total Hamiltonian $H^{\rm osc}_0 = \hbar \omega_0 a^{\dagger}_{1}  a_{1} \otimes 1_2
%+ 1_1 \otimes \hbar \omega_0 a^{\dagger}_{2}  a_{2} $, 
As before, we map the CF on the Keldysh contour in the interaction picture with respect to the bare part of the total Hamiltonian $H^{\rm osc}_0 = \hbar \omega_0 a^{\dagger}_{1}  a_{1} \otimes 1_2
+ 1_1 \otimes \hbar \omega_0 a^{\dagger}_{2}  a_{2} $ as
\be
\chi_{T}(u) = {\rm Tr} \Big[ \rho(0) T_c e^{-i \int_c V^x(\tau) \, d\tau}  \Big],
\label{contour-eq}
\ee
where the interaction Hamiltonian in this case is dressed as $V^x = e^{i x H_1} V e^{-i x H_1} = J ({a}_1^{x\, \dagger} \, a_2 + h.c.)$, $x= \pm u/2$, affecting only the system 1 operators. Note that, the operator $V^x$ is time-independent even in the interaction picture due to the commutable coupling symmetry. However in Eq.~(\ref{contour-eq}) we  explicitly write the contour-ordered operator to keep track of the forward and the backward evolution. Recall that the contour time variable $\tau$ runs from $[0,T]$.
%Expanding the exponential in Eq.~(\ref{contour-eq}) and collecting terms order-by-order for the coupling $J$ and finally 
Invoking the linked-cluster theorem for the CGF ${\cal G}_{T}^{\rm osc}(u) = \ln \chi_{T}(u)$ we receive a formal exact expression for the model in contour-time $\tau$  as
\be
{\cal G}_{T}^{\rm osc}(u)= -{\rm Tr}_{\tau} \ln \Big[1- g_{22} \, \Sigma_{11}^{x} \Big].
\label{exact-CGF}
\ee
Here the Green functions are understood as matrices in discretized contour time. In continuous time version the trace operation means ${\rm Tr}_{\tau}[A\, B] = \int d\tau \int d\tau' A(\tau,\tau') B(\tau',\tau)$. In the above expression, following the standard notations for the Green functions, we define 
\begin{align}
g_{ii}(\tau,\tau') &= -i \langle T_c a_i (\tau) \, a_i^{\dagger} (\tau) \rangle, \quad i=1,2 \label{bare} \\
g^{x}_{ii}(\tau,\tau') &=  -i \langle T_c {a}^{x}_i(\tau) \, {a}^{x'\dagger}_i (\tau') \rangle, \quad i=1
\label{dressed}
\end{align}
as the bare (Eq.~\ref{bare}) and the counting field dependent (Eq.~\ref{dressed}) Greens function, respectively. Recall that, the counting field appears only for system 1 operators. The self-energy term is then given as $\Sigma^{x}_{11}(\tau,\tau') = J^2 g^{x}_{11}(\tau,\tau')$ with $J$ being the coupling strength between the oscillators. %where $g^{x}_{11}(\tau,\tau') =  -\frac{i}{\hbar} \langle T_c {a}_1(\tau)^{x(\tau)} \, \tilde{a}_1^{x(\tau') \dagger} (\tau') \rangle$.  
Since $a_1^{x}(\tau) = e^{i x H_1} a_1(\tau) e^{-i x H_1} = a_1(\tau+ x(\tau))$,  it is thus clear that the effect of measuring or counting energy leads to a shift in contour-time and correspondingly the self-energy is shifted as
\be
\Sigma_{11}^{x}(\tau,\tau') = \Sigma_{11}(\tau + x (\tau), \tau' + x(\tau')).
\label{shift}
\ee
Eq.~(\ref{exact-CGF}) doesn't explicitly satisfy the normalization conduction ${\cal G}_{T}^{\rm osc}(u=0)=0$. To enforce this condition, one can further simplify the above expression and write 
\bea
1- g_{22} \Sigma^{\rm x}_{11} &=& g_{22} \Big(g^{-1}_{22}- \Sigma^{\rm x}_{11}\Big) \nonumber \\
&& = g_{22} \Big(g^{-1}_{22}- \Sigma- \Sigma^{\rm A}_{11}\Big) \nonumber \\
&& =g_{22} \Big(G^{-1}_{22}- \Sigma^{\rm A}_{11}\Big) \nonumber \\
&& = g_{22}\, G^{-1}_{22} \Big(1- G_{22}\,\Sigma^{\rm A}_{11}\Big)  \nonumber \\
&&= \Big(1- g_{22} \Sigma_{11}\Big) \Big(1- G_{22}\,\Sigma^{\rm A}_{11}\Big),
\label{norm}
\eea
where in the second line we define a useful quantity $\Sigma_{11}^A =\Sigma^{x}_{11}- \Sigma_{11}$ which is zero in the absence of the counting field. The third line motivates one to introduce a new Green's function $G^{-1}_{22} = g^{-1}_{22}- \Sigma_{11}$ which in continuous contour-time version satisfies the following Dyson equation:
\bea
G_{22}(\tau,\tau') &=& g_{22}(\tau,\tau') + \int_{c} \int_{c} d\tau_1 d\tau_2  \, g_{22}(\tau, \tau_1) \Sigma_{11}(\tau_1,\tau_2) \nonumber \\
&&\, G_{22}(\tau_2,\tau').
\eea
Notice that, this Green's function is nothing but the dressed Greens function of system 2, taking into account the presence of system 1 in terms of the self-energy $\Sigma_{11}$.  With the help of Eq.~(\ref{norm}), Eq.~(\ref{exact-CGF}) then simplifies to 
\bea
{\cal G}_{T}^{\rm osc}(u)= -{\rm Tr}_{\tau} \ln \Big[1- G_{22}  \, \Sigma_{11}^{A} \Big]
\eea
as ${\rm Tr}_{\tau} \ln \Big[1- g_{22}  \Sigma_{11} \Big]=0$ following Eq.~(\ref{exact-CGF}), ensuring the normalization condition.
%The condition ${\cal G}_{\tau}^{\rm osc}(u=0) =0$ ensures that

The next important task from here on is to go from the contour-time to the real time following the Langreth’s theorem. Furthermore, a more transparent and simplified framework is obtained by performing an orthogonal Keldysh rotation (rotation in the space of real time by {45}\degree) which gives
\bea
{\cal G}_{T}^{\rm osc}(u)= -{\rm Tr}_{t, \sigma} \ln \Big[1- \breve{G}_{22}  \, \breve{\Sigma}_{11}^{A} \Big].
\label{real-CGF}
\eea
The breve symbol indicates that the Green's functions are written in the rotated Keldysh frame. Also note that, the orthogonal Keldysh rotation preserves the trace in the above CGF expressions. In Eq.~(\ref{real-CGF}) the meaning of trace is now in terms of the real time and as well as over the branch index, denoted as $\sigma$. Explicitly,it means, for example, ${\rm Tr}_{t,\sigma}[\breve{A}\, \breve{B}] = \int_{0}^{T} dt_1 \int_{0}^{T} dt_2 {\rm Tr}\big[\breve{A}(t_1,t_2) \breve{B}(t_2,t_1)\big]$. We receive the $\breve{G}_{22}$ as,
%Here the breve symbol on the Green's function imply that the matrices are obtained after the Keldysh rotation (denoted by the breve symbol) are given as
%\ee the subscript $t$ indicates that the above trace is now in real time and $\sigma$, the branch index corresponds to trace over the real-time matrix. Explicity, it means
%The symbol where in tHere the counting field dependence appears in $\breve{\Sigma}_{11}^A(\omega_0)$ and the counting field independent Green's function matrix $\breve{G}_{22}$ is given as, 
\begin{equation}
 \breve{G}_{22} =
\begin{bmatrix}
G_{22}^r & G_{22}^k  \\
0 & G_{22}^a \\
\end{bmatrix},
\end{equation}
where ${r,a,k}$ are the retarded, advanced and the Keldysh components for the Green function. These various components can be obtained exactly and are given as follows:
\begin{eqnarray}
G_{22}^r (t,t') &=& -i  \, \theta(t-t') e^{-i \omega_0 (t-t')} \, \cos(J (t-t')), \nonumber \\
G_{22}^a (t,t') &=& i \, \theta(t'-t) e^{-i \omega_0 (t-t')} \, \cos(J (t-t')),\nonumber  \\
G_{22}^< (t,t') & = & -i \, \Big[ n_2 \, \cos(Jt) \, \cos (Jt') + n_1 \, \sin (Jt) \, \sin (Jt') \Big],\nonumber \\
G_{22}^> (t,t') & \!= & -i \, \Big[ (1\!+ n_2) \cos(J t) \, \cos (Jt') +(1\!+ n_1) \nonumber \\
&& \quad \quad \!\, \sin (Jt) \, \sin (Jt')\Big], \nonumber \\
\end{eqnarray}
and the Keldysh component is given as $ G_{22}^{k} = G_{22}^{<} + G_{22}^{>}$. Interestingly, the retarded and the advanced components are time-translational invariant which is not the case for other components. It is easy to check that the lesser and greater components satisfy the correct initial condition, given as $ i \, G_{22}^{<}(t\!=\!t'=\!0) = \langle a_{2}^{\dagger} a_2 \rangle =  \, n_2$ and $i\, G_{22}^{>}(t=t'=0) =  \langle a_{2} a_2^{\dagger} \rangle = (1+ n_2)$. 
%Note that for notational compactness we have suppress the arguments in both $\tilde{G}$ and $\Sigma$. 
Similarly we receive for the counting field dependent self-energy as
\begin{equation}
\breve{\Sigma}_{11}^A= \frac{1}{2} 
\begin{bmatrix}
a-b & a+b  \\
-(a+b) &b-a \\
\end{bmatrix},
\end{equation}
where
\begin{align}
a&=  \Sigma_{11}^{>}(t-t'+ u)\! -\! \Sigma_{11}^{>}(t,t'),  \nonumber \\
b&= \Sigma_{11}^{<}(t-t'-u) \!-\! \Sigma_{11}^{<}(t,t').
\end{align}
The calculation further simplifies upon performing a two-time Fourier transformation, defined here as
\be
\breve{G}_{22}(\omega_1, \omega_2) = \int_{0}^{T} dt \, \int_{0}^{T} dt' \, \, e^{i \omega_1 t} \, e^{i \omega_2 t'} \,\, \breve{G}_{22}(t,t').
\label{Fourier}
\ee
One then finally obtains from Eq.~(\ref{real-CGF})
\begin{equation}
{\cal G}_{T}^{\rm osc} (u) = - {\rm ln} \, {\rm det} \Big[ 1 - \breve{G }_{22} (\omega_0, -\omega_0) \breve{\Sigma}_{11}^A(\omega_0)\Big].
\label{freq-CGF}
\end{equation}
%where we define the two component Fourier transformation as 
Note that, the above formula is exact for arbitrary coupling $J$. This expression can be easily extended for many-oscillator setup also. One can now write down the  Fourier version of the Green functions components which are given as,
%In the frequency domain, the various correlators are given as 
\bea
G_{22}^r (\omega_0, -\omega_0) &=& -\frac{2 i}{ J^2 }\, \sin^2 \big(\frac{J T}{2}\big),\nonumber \\
G_{22}^a (\omega_0, -\omega_0) &=& \frac{2 i}{ J^2} \, \sin^2 \big(\frac{J T}{2}\big),\nonumber \\
G_{22}^< (\omega_0, -\omega_0) &=& -\frac{i}{ J^2 } \Big[ n_2   \sin^2 (J T) + n_1 \big(1- \cos(J T)\big)^2 \big],\nonumber \\
G_{22}^> (\omega_0, -\omega_0) &=& -\frac{i}{ J^2 } \Big[(1+ n_2)   \sin^2 (J \,T) + (1+ n_1) \nonumber \\
&& \big(1- \cos(J\,T)\big)^2 \big],\nonumber \\
\eea
and similarly for the self-energy components,
\begin{eqnarray}
a &=& \Sigma_{11}^{>}(\omega_0) \, (e^{-i u \hbar \omega_0} -1)  \nonumber \\
&=& -i\,J^2\,  \big( 1+ n_1(\omega_0)\big) \, (e^{-i u \hbar \omega_0} -1), \\ 
b &=& \Sigma_{11}^{<}(\omega_0)\,  (e^{i u \hbar \omega_0} -1) \nonumber \\
&=&  -i \, J^2 \, n_1(\omega_0) \,  (e^{i u \hbar \omega_0} -1).
\end{eqnarray}
Knowing these analytical expressions for the Green functions one can  simply compute the determinant in Eq.~(\ref{freq-CGF}), which finally gives the CGF expression in Eq.~(\ref{CGF:osc}). 
%\begin{bmatrix}
%\begin{equation}
%\!\!\!\!\!\!\chi_{\tau}(u)\!\!=\!\! \langle U_{I}^{\dagger}\big]^{-\xi/2}(t,0) \, U_{I}^{\xi/2}(t,0) \rho(0) \Big]
%\end{equation}

%\begin{equation}
%\!\!\!\!\!\!\chi_{\tau}(u)\!\!=\!\! {\rm Tr}\Big[\big[U_{I}^{\dagger}\big]^{-\xi/2}(t,0) \, U_{I}^{\xi/2}(t,0) \rho(0) \Big]
%\end{equation}

\begin{comment}
\begin{figure}
\centering
\includegraphics[trim= 0cm 0cm 0cm 0cm, clip=true,width=9cm]{JC_b2=1_Jt.pdf} 
\caption{{ \textcolor{blue}{Jaynes-Cummings Model: First three cumulants of heat exchange, along with a measure for the S-TUR,
as a function of the coupling strength $J\tau $; $\beta_1=0$; $\beta_2=1$.
The solid lines are constructed with the help of Eq. (\ref{cumulants:JC}) and the summation is taken upto $n=30$ in Eq. (\ref{eq: sum_QQ})}} }
\label{JC_TUR_Cumulants}
\end{figure}
\end{comment}

%The expression for CGF in the weak-coupling limit was previously obtained in Ref.~(). 

%=======================================

%====================================================

\begin{thebibliography}{999}
%=============================================


\bibitem{Ritort} F. Ritort, Nonequilibrium fluctuations in small systems:
from physics to biology,  Adv. Chem. Phys. {\bf 137} 31–123 (2008).

\bibitem{spin-heat} J. P. S. Peterson et.al., Experimental Characterization of a spin quantum heat engine,  Phys. Rev. Lett. {\bf 123}, 240601 (2019).

\bibitem{heat-fluc} S. Rahav, U. Harbola, and S. Mukamel, Heat fluctuations and coherences in a quantum heat engine, Phys. Rev. A {\bf 86}, 043843 (2012).

%-----------------FLUCTUTATION--------------------------
%================================

\bibitem{fluc-old1} D. J. Evans, E. G. D. Cohen and G. P. Morriss, Probability of second law violations in shearing steady states, Phys. Rev. Lett. {\bf 71}, 2401 (1993).

\bibitem{fluc-old2} G. Gallavotti and E. G. D. Cohen  Dynamical ensembles in nonequilibrium statistical mechanics, Phys. Rev. Lett.
{\bf 74}, 2694 (1995).

\bibitem{fluc-old2a} C. Jarzynski, Nonequilibrium equality for free energy differences Phys. Rev. Lett. {\bf 78} 2690 (1997). 

\bibitem{fluc-old3} J. Kurchan, Fluctuation theorem for stochastic dynamics J. Phys. A: Math. Gen. {\bf 31}, 3719 (1998).

\bibitem{fluc-old4} J. L. Lebowitz and H. Spohn, A Gallavotti-Cohen-type symmetry in the large deviation functional for stochastic
dynamics, J. Stat. Phys. {\bf 95}, 333 (1999).


\bibitem{JarzW}
C. Jarzynski and D. K. Wójcik,
Classical and quantum fluctuation theorems for heat exchange,
Phys. Rev. Lett. {\bf 92}, 230602 (2004).
% https://journals.aps.org/prl/abstract/10.1103/PhysRevLett.92.230602




\bibitem{fluct1}
M. Esposito, U. Harbola, and S. Mukamel,
Nonequilibrium fluctuations, fluctuation theorems, and counting statistics in quantum systems,
Rev. Mod. Phys. {\bf 81}, 1665 (2009).

\bibitem{fluct2}
M. Campisi, P. H\"anggi, and  P. Talkner,
Colloquium: Quantum fluctuation relations: Foundations and applications,
Rev. Mod. Phys. {\bf 83}, 771 (2011).


\bibitem{fluct3} C. Jarzynski, Equalities and inequalities: irreversibility and the second law of thermodynamics at the nanoscale, Annu. Rev. Condens. Matter Phys. {\bf 2} 329 (2011).



\bibitem{SaitoUts}
K. Saito and Y. Utsumi,
Symmetry in full counting statistics, fluctuation theorem, and relations among nonlinear transport coefficients in the presence of a magnetic field,
Phys. Rev. B \textbf{78},  115429 (2008).


\bibitem{campisi-measurement}
M. Campisi, P. Talkner, and P. H\"anggi,
Influence of measurements on the statistics of work performed on a quantum system,
Phys. Rev. E {\bf 83}, 041114 (2011).


%------------------------------------------



\bibitem{st-thermo1} 
U. Seifert,  Stochastic thermodynamics, fluctuation theorems, and molecular machines, Rep. Prog. Phys. {\bf 75}, 126001 (2012).


\bibitem{st-thermo2}
U. Seifert, 
Stochastic thermodynamics: Principles and perspective, Eur. Phys. J. B, {\bf 64}, 423 (2008).
 
\bibitem{QT1} R. Kosloff, Quantum thermodynamics and open-systems modeling,
J. Chem. Phys., {\bf 150}, 204105,  (2019)


\bibitem{QT2} 
S. Vinjanampathy and J. Anders, 
Quantum thermodynamics, Contemporary Physics, {\bf 57}, 545 (2016).

%-------------------------------------
%-----------------------TUR----------------------

\bibitem{Barato:2015:UncRel}
A. C. Barato and U. Seifert,
Thermodynamic uncertainty relation for biomolecular processes,
Phys. Rev. Lett. {\bf 114}, 158101 (2015).
% https://journals.aps.org/prl/abstract/10.1103/PhysRevLett.114.158101

\bibitem{Gingrich:2016:TUP}
T. R. Gingrich, J. M.  Horowitz, N. Perunov, and J. L. England,
Dissipation bounds all steady state current fluctuations,
Phys. Rev. Lett. {\bf 116}, 120601 (2016).


\bibitem{Polettini:2016:TUP}
M. Polettini, A. Lazarescu, and M. Esposito, 
Tightening the uncertainty principle for stochastic currents, 
Phys. Rev. E {\bf 94}, 052104 (2016).

\bibitem{Pietzonka:2016:Bound}
P. Pietzonka, A. C. Barato, and U. Seifert, 
Universal bounds on current fluctuations,
Phys. Rev. E {\bf 93}, 052145 (2016).
% https://journals.aps.org/pre/abstract/10.1103/PhysRevE.93.052145

\bibitem{Hyeon:2017:TUR}
C. Hyeon and W. Hwang, 
Physical insight into the thermodynamic uncertainty relation using Brownian motion in tilted periodic potentials,
Phys. Rev. E {\bf 96}, 012156 (2017).

\bibitem{Horowitz:2017:TUR} 
J. M. Horowitz and T. R. Gingrich, 
Proof of the finite-time thermodynamic uncertainty relation for steady-state currents, 
Phys. Rev. E {\bf 96}, 020103(R) (2017).

\bibitem{Pigolotti:TURF}
S. Pigolotti, I. Neri, E. Roldán, and F. J\"ulicher, 
Generic Properties of Stochastic Entropy Production,
Phys. Rev. Lett. {\bf 119}, 140604 (2017).

\bibitem{Proesmans:2017:TUR}
K. Proesmans and C. V. den Broeck, 
Discrete-time thermodynamic uncertainty relation,
EPL {\bf 119}, 20001 (2017).


\bibitem{Hwang}
W. Hwang, and C. Hyeon,
Energetic costs, precision, and transport efficiency of molecular motors,
J. Phys. Chem. Lett. {\bf 9}, 513 (2018).
% 



\bibitem{Bio}
R. Marsland III, W. Cui and J. M. Horowitz,
The thermodynamic uncertainty relation in biochemical oscillations,
J.  R.  Soc.  Interface {\bf 16}, (2019).  % DDD?
% https://doi.org/10.1098/rsif.2019.0098

\bibitem{Interacting}
S. Lee, C. Hyeon, and J. Jo, 
Thermodynamic uncertainty relation of interacting oscillators in synchrony,
Phys. Rev. E {\bf 98}, 032119 (2018).
% https://journals.aps.org/pre/abstract/10.1103/PhysRevE.98.032119


\bibitem{Mayank}
M. Shreshtha and R. J. Harris,
Thermodynamic uncertainty for run-and-tumble-type processes,
EPL {\bf 126}, 40007  (2019). 
%https://iopscience.iop.org/article/10.1209/0295-5075/126/40007

\bibitem{Garrahan:2017:TUR}
J. P. Garrahan, 
Simple bounds on fluctuations and uncertainty relations for first-passage times of counting observables,
Phys. Rev. E {\bf 95}, 032134 (2017).
% https://journals.aps.org/pre/abstract/10.1103/PhysRevE.95.032134


\bibitem{Passage}
T. R. Gingrich and J. M. Horowitz,
Fundamental bounds on first passage time fluctuations for currents,
Phys. Rev. Lett. {\bf 119}, 170601 (2017).

\bibitem{Dechant:2018:TUR}
A. Dechant, 
Multidimensional thermodynamic uncertainty relations,
J. Phys. A: Math. Theor. {\bf 52}, 035001 (2019).
% https://iopscience.iop.org/article/10.1088/1751-8121/aaf3ff

%\bibitem{TUR5} 
%P. Pietzonka and U. Seifert, 
%Universal Trade-Off between Power, Efficiency, and Constancy in Steady-State Heat Engines, 
%Phys. Rev. Lett. {\bf 120}, 190602 (2018).

\bibitem{Pietzonka:2017:FiniteTUR} 
P. Pietzonka, F. Ritort, and U. Seifert,
Finite-time generalization of the thermodynamic uncertainty relation, 
Phys. Rev. E {\bf 96}, 012101 (2017).

\bibitem{Falasco}
G. Falasco, M. Esposito, and J.-C. Delvenne,
Unifying thermodynamic uncertainty relations,
arXiv:1906.11360.
% https://arxiv.org/abs/1906.11360

%Generalized TUR:
\bibitem{SamuelssonM}
P. P. Potts and P. Samuelsson,
Thermodynamic uncertainty relations including measurement and feedback,
Phys. Rev. E {\bf 100}, 052137 (2019).

\bibitem{Koyuk:2018:PeriodicTUR}
T. Koyuk, U. Seifert, and P. Pietzonka, 
A generalization of the thermodynamic uncertainty relation to periodically driven systems,
J. Phys. A: Math. Theor. {\bf 52}, 02LT02 (2018).
% https://iopscience.iop.org/article/10.1088/1751-8121/aaeec4/meta

\bibitem{Garrahan18}
K. Macieszczak, K. Brandner, and J. P. Garrahan, 
Unified thermodynamic uncertainty relations in linear response,
Phys. Rev. Lett. {\bf 121}, 130601 (2018).
% https://journals.aps.org/prl/abstract/10.1103/PhysRevLett.121.130601

\bibitem{Van}
T. Van Vu  and  Y. Hasegawa.
Uncertainty  relation  in  the  presence  of  information measurement and feedback control,
 arXiv:1904.04111


\bibitem{Hyst}
K. Proesmans and J. M. Horowitz,
Hysteretic thermodynamic uncertainty relation for systems with broken time-reversal symmetry,
J. Stat. Mech. 054005 (2019).

\bibitem{Gabri}
A. C. Barato, R. Chetrite, A. Faggionato, and D. Gabrielli, 
Bounds on current fluctuations in periodically driven systems,
New J. Phys. {\bf 20}, 103023 (2018).
 

\bibitem{Vu}
Y. Hasegawa and T. Van Vu,
Fluctuation theorem uncertainty relation
Phys. Rev. Lett. {\bf 123}, 110602 (2019).
%Generalized thermodynamic uncertainty relation via the fluctuation theorem,
%arXiv:1902.06376v3.
%https://arxiv.org/pdf/1902.06376.pdf

\bibitem{Hasegawa1} 
Y. Hasegawa and T. V. Vu, 
Uncertainty relations in stochastic processes: An information inequality approach, 
Phys. Rev. E {\bf 99}, 062126  (2019). %DDD year
%https://arxiv.org/abs/1809.06610


\bibitem{Hasegawa2} 
M. L. Rosinberg and G. Tarjus, 
Comment on Thermodynamic uncertainty relation for time-delayed Langevin systems, 
arXiv:1810.12467.
%https://arxiv.org/pdf/1810.12467.pdf

\bibitem{Gingrich:2017}
T. R. Gingrich, G. M. Rotskoff and J. M Horowitz,
Inferring dissipation from current fluctuations, 
J.Phys. A: Math. Theor. {\bf 50} 184004 (2017). % DDD check

\bibitem{Sasa:TUR}
A. Dechant and S-i. Sasa, 
Current fluctuations and transport efficiency for general Langevin systems,
J. Stat. Mech. 063209 (2018). 
%https://iopscience.iop.org/article/10.1088/1742-5468/aac91a

\bibitem{Baiesi}
I. Di Terlizzi and M. Baiesi, 
Kinetic uncertainty relation, 
J. of Phys. A: Math. and Theor. {\bf 52}, 02LT03 (2018).
% https://iopscience.iop.org/article/10.1088/1751-8121/aaee34/meta

\bibitem{Saito}
K. Brandner, T. Hanazato, and K. Saito,
Thermodynamic bounds on precision in ballistic multiterminal transport,
Phys. Rev. Lett. {\bf 120}, 090601 (2018).

\bibitem{TUR-gupta}
D. Gupta and A. Maritan,
Thermodynamic uncertainty relations via second law of thermodynamics,
Eur. Phys. J. B {\bf 93}, 28 (2020).
%arXiv:1905.08854.

\bibitem{Udo:TURB}
H.-M. Chun, L. P. Fischer, and U. Seifert,
Effect of a magnetic field on the thermodynamic uncertainty relation, 
Phys. Rev. E {\bf 99}, 042128 (2019).

\bibitem{TURQ}
K. Ptaszynski,
Coherence-enhanced constancy of a quantum thermoelectric generator,
Phys. Rev. B {\bf 98}, 085425 (2018). %Nat. Phys. (2019) 
%https://journals.aps.org/prb/abstract/10.1103/PhysRevB.98.085425
% DDD?

\bibitem{BijayTUR}
B. K. Agarwalla and D. Segal,
Assessing the validity of the thermodynamic uncertainty relation in quantum systems,
Phys. Rev. B {\bf 98}, 155438 (2018).

\bibitem{TUR-bijay1}
S. Saryal, H. Friedman, D. Segal, and B. K. Agarwalla, 
Thermodynamic uncertainty relation in thermal transport, 
Phys. Rev. E {\bf 100}, 042101 (2019).


\bibitem{JunjieTUR}
J. Liu and D. Segal,
Thermodynamic uncertainty relation in quantum thermoelectric junctions,
Phys. Rev. E {\bf 99}, 062141 (2019).

\bibitem{SamuelssonQP}
S. Kheradsoud, N. Dashti, M. Misiorny, P. P. Potts, J. Splettstoesser, and P. Samuelsson,
Power, efficiency and fluctuations in a quantum point contact as steady-state thermoelectric heat engine,
Entropy {\bf 21}, 777 (2019).
%https://www.mdpi.com/1099-4300/21/8/777





\bibitem{Landi-PRL} 
A. M. Timpanaro, G. Guarnieri, J. Goold, and G. T. Landi, 
Thermodynamic uncertainty relations from exchange fluctuation theorems, 
Phys. Rev. Lett. {\bf 123}, 090604 (2019).

\bibitem{Isometric}
H. Vroylandt, K. Proesmans, and T. R. Gingrich,
Isometric Uncertainty Relations
arXiv:1910.01086.


%\bibitem{VanTUR}
%Y. Hasegawa and T. Van Vu,
%Fluctuation theorem uncertainty relation,
%Phys. Rev. Lett. {\bf 123}, 110602 (2019). % DDD check

\bibitem{Goold}
G. Guarnieri, G. T. Landi, S. R. Clark, and J. Goold,
Thermodynamics of precision in quantum non-equilibrium steady states,
Phys. Rev. Research {\bf 1}, 033021 (2019).
%arXiv:1901.10428.

\bibitem{Horowitz:2019:TUR}
J. M. Horowitz and T. R. Gingrich
Thermodynamic uncertainty relations constrain non-equilibrium fluctuations,
Nat. Phys. (2019).
%https://doi.org/10.1038/s41567-019-0702-6


\bibitem{TUR-driving-udo} T. Koyuk and U. Seifert, Thermodynamic uncertainty relation for time-dependent driving, arXiv: 2005.02312.

\bibitem{TUR-opensys} Y. Hasegawa, Thermodynamic uncertainty relation for open quantum systems, arXiv: 2003.08557.

\bibitem{TUR-Otto} M. F. Sacchi, Thermodynamic Uncertainty Relations for Bosonic Otto Engines, arXiv: 2007.05399.



\bibitem{tur_energy}
M. W. Jack, N. J. López-Alamilla, and K. J. Challis, Thermodynamic uncertainty relations and molecular-scale energy conversion. Phys. Rev. E {\bf 101}, 062123 (2020).

\bibitem{tur_superconductor}
H. Tajima, K. Funo, Superconducting-like heat current:Effective cancellation of current-dissipation trade off by quantum coherence.
arXiv:2004.13412.

\bibitem{tur_turing}
S. Rana, A. C. Barato, Precision and dissipation of a stochastic Turing pattern, arXiv:2004.01230

\bibitem{tur_hyper}
D. M. Busiello and S. Pigolotti,
Hyperaccurate currents in stochastic thermodynamics, Phys. Rev. E {\bf 100}, 060102(R) (2019).

\bibitem{tur_continuous}
Y. Hasegawa,
Quantum Thermodynamic Uncertainty Relation for Continuous Measurement, Phys. Rev. Lett {\bf 125}, 050601 (2020).



\bibitem{Soham_TUR}  S. Pal, S. Saryal, D. Segal, T. S. Mahesh, and B. K. Agarwalla, Experimental study of the thermodynamic uncertainty relation,  Phys. Rev. Research (R) {\bf 2}, 022044 (2020).


\bibitem{Oren}
H. M. Friedman, B. K. Agarwalla, O. Shein-Lumbroso, O. Tal, and D. Segal,
Thermodynamic uncertainty relation in atomic-scale quantum conductors,Phys. Rev. B, {\bf 101}, 195423 (2020).


%Experimental Papers
\bibitem{tur_kproof}
W. D. Pineros and T. Tlusty,
Kinetic proofreading and the limits of thermodynamic uncertainty, Phys. Rev. E {\bf 101}, 022415 (2020).

\bibitem{tur_infoengine}
G. Paneru, S. Dutta, T. Tlusty, H. K. Pak,
Approaching and violating thermodynamic uncertainty bounds in measurements of fluctuation-dissipation tradeoffs of information engines. arXiv:1911.09835

%Theory papers


\bibitem{Supriya} S. K. Manikandan, D. Gupta, and S. Krishnamurthy, Inferring Entropy Production from Short Experiments,
Phys. Rev. Lett. {\bf 124}, 120603 (2020).

\bibitem{XFT-agarwalla}
B. K. Agarwalla, H. Li, B. Li, and J.-S. Wang,
Exchange fluctuation theorem for heat transport between multiterminal harmonic systems,
Phys. Rev. E {\bf 89}, 052101 (2014).

\bibitem{Bijay12}
B. K. Agarwalla, B. Li, and  J.-S. Wang,
Full-counting statistics of heat transport in harmonic junctions: Transient, steady states, and fluctuation theorems,
Phys. Rev. E {\bf 85}, 051142 (2012).

\bibitem{Saito07}
K. Saito and A. Dhar,
Fluctuation theorem in quantum heat conduction,
Phys. Rev. Lett. \textbf{99}, 180601 (2007).

\bibitem{Lutz_2018}
T. Denzler and E. Lutz,
Heat distribution of a quantum harmonic oscillator,
Phys. Rev. E {\bf 98}, 052106  (2018).

\bibitem{XFT-theory}
G. T. Landi and D. Karevski,
Fluctuations of the heat exchanged between two quantum spin chains,
Phys. Rev. E {\bf 93}, 032122 (2016).



\bibitem{NEGF1} J. Schwinger, Brownian Motion of a Quantum Oscillator, J. Math. Phys. {\bf 2}, 407 (1961).
\bibitem{NEGF2} L. V. Keldysh, Diagram technique for nonequilibrium processes, Sov. Phys.  JETP {\bf 20}, 1018 (1965).
\bibitem{NEGF3} J. Rammer and H. Smith, Quantum field-theoretical methods in transport theory of metals, Rev. Mod. Phys. {\bf 58}, 323 (1986).
\bibitem{Mahan} G. D. Mahan, {\it Many-Particle Physics}, 3rd ed. (Kluwer Academic, New York, 2000).
\bibitem{NEGF4} H. Haug and A.-P. Jauho, {\it Quantum Kinetics in Transport and
Optics of Semiconductors}, 2nd ed. (Springer, New York, 2008).


\bibitem{XFT-Akagawa} S. Akagawa and N. Hatano, The Exchange Fluctuation Theorem in Quantum Mechanics, Prog. Theor. Phys. {\bf 121}, 1157 (2009).


%\bibitem{Markov-Rev}
%L. Li, M. J. Hall, and H. M. Wiseman, 
%Concepts of quantum non-Markovianity: A hierarchy,
%Phys. Rep. {\bf 759}, 1 (2018).




%\bibitem{Wei} B. B. Wei,
%Relations between heat exchange and R\'enyi divergences,
%Phys. Rev. E {\bf 97}, 042107 (2018).


%==========================================
%\bibitem{supp}
%See supplemental Material File
%==========================================


%%ancilla proposal
\bibitem{ancilla-1}
L. Mazzola, G. De Chiara, and M. Paternostro,
Measuring the characteristic function of the work distribution,
Phys. Rev. Lett. {\bf 110}, 230602 (2013).
%proposals-ancilla

\bibitem{ancilla-2}
R. Dorner, S. R. Clark, L. Heaney, R. Fazio, J. Goold, and V. Vedral,
Extracting quantum work statistics and fluctuation theorems by single-qubit interferometry,
Phys. Rev. Lett. {\bf 110}, 230601  (2013).
%proposals ancillaa


\bibitem{ancilla-3}
M. Campisi, R. Blattmann, S. Kohler, D. Zueco, and P. Hanggi,
Employing circuit QED to measure non-equilibrium work fluctuations,
New J. Phys. {\bf 15}, 105028 (2013).


%============================= 

\bibitem{Bijay-expt} 
S. Pal, T. S. Mahesh, and B. K. Agarwalla, 
Experimental demonstration of the validity of the quantum heat-exchange fluctuation relation in an NMR setup,
 Phys. Rev. A {\bf 100}, 042119 (2019).

%=============================



%\bibitem{Tal_TUR} H. Friedman, B. K. Agarwalla, O. S.-Lumbroso, O.Tal, and D. Segal, Thermodynamic uncertainty relation in atomic-scale quantum conductors,
%Phys. Rev. B, {\bf 101}, 195423 (2020).





\bibitem{Gerry} C. Gerry and  P. Knight,  Introductory Quantum Optics. University Press, Cambridge (2005).

%\bibitem{comment}
%In the weak coupling limit the second cumulant (\ref{eq:analM}) can be organized as (using $x \coth(x) \geq 1$)
%\bea
%\langle Q^2 \rangle^{c}_{\tau} &=& (h \nu_0)^2 {\cal T}_{\tau}(J) \Big( f_1 (1\!-\!f_2) \!+\! f_2 (1\!-\!f_1)\Big) \nonumber \\
%&=&  (h \nu_0)^2  \coth\Big(\frac{\Delta \beta \, h \nu_0}{2}\Big) {\cal T}_{\tau}(J) (f_2 - f_1) \nonumber \\
%&\geq & \frac{2}{\Delta \beta} h \nu_0 {\cal T}_{\tau}(J) (f_2 - f_1) = \frac{2}{\Delta \beta} \langle Q \rangle_{\tau},
%\eea
%
%proving that the S-TUR is satisfied even far from equilibrium.

%\bibitem{heat-distribution} 
%J. P. S. Peterson, T. B. Batalh\"ao, M. Herrera, A. M. Souza, R. S. Sarthour, I. S. Oliveira, and R. M. Serra,
%Experimental characterization of a spin quantum heat engine, 
%arXiv:1803.06021.
%Phys. Rev. Lett. {\bf 123}, 240601  (2019).


\end{thebibliography}
\end{document}